\documentclass[aps,prd,twocolumn,preprintnumbers,showpacs,superscriptaddress,nofootinbib,floatfix,10pt]{revtex4-2}
\usepackage{textcomp}
\usepackage[english]{babel}
\usepackage{amsmath,amssymb,amsbsy,booktabs}
\usepackage{bm}
\usepackage{color}
\usepackage{array}
\usepackage{amstext}
\usepackage{graphicx}
\usepackage{amsfonts}
\usepackage{dcolumn}
\usepackage{rotating}
\usepackage{epstopdf}
\usepackage{esint}
\usepackage{url}
\usepackage[colorlinks=true,
linkcolor=blue,
filecolor=blue,
anchorcolor=blue,
urlcolor=blue,
citecolor=blue
]{hyperref}

\usepackage{array}
\usepackage{booktabs}
\usepackage{multirow}

\usepackage[utf8]{inputenc}

\usepackage{xcolor,colortbl}

\usepackage[T1]{fontenc} 

\usepackage{cancel}
\usepackage{slashed}
\usepackage{hhline}
\usepackage{subfigure} 
\usepackage{enumerate}
\usepackage{times}
\usepackage[version=4]{mhchem}

\DeclareMathOperator{\tr}{Tr}

\allowdisplaybreaks

\begin{document}

\title{\Large \bf \boldmath Polarization Probes of New Physics in Lepton-Flavor-Violating Hyperon Production from $e^- N \to \tau^- Y$ Scattering}

\author{Xin-Shuai Yan}
\email{yanxinshuai@htu.edu.cn}
\affiliation{Institute of Particle and Nuclear Physics, Henan Normal University, Xinxiang, Henan 453007, China}

\author{Zhen-Qing Hu}
\email{huzhenqing@stu.htu.edu.cn}
\affiliation{Institute of Particle and Nuclear Physics, Henan Normal University, Xinxiang, Henan 453007, China}

\author{Qin Chang}
\email{changqin@htu.edu.cn (corresponding author)}
\affiliation{Institute of Particle and Nuclear Physics, Henan Normal University, Xinxiang, Henan 453007, China}

\author{Ya-Dong Yang}
\email{yangyd@mail.ccnu.edu.cn}
\affiliation{Institute of Particle and Nuclear Physics, Henan Normal University, Xinxiang, Henan 453007, China}
\affiliation{Institute of Particle Physics and Key Laboratory of Quark and Lepton Physics~(MOE),\\
	Central China Normal University, Wuhan, Hubei 430079, China}

\begin{abstract}

While charged lepton-flavor-violating tau decays to strange mesons provide powerful probes of the underlying 
$\tau\to\ell d \bar{s}$ transition, the corresponding decay modes involving hyperons are kinematically forbidden. 
To address this gap, we propose the quasi-elastic scattering processes $e^- N \to \tau^- Y$. 
Within a general low-energy effective Lagrangian, 
we perform a comprehensive analysis of the final-state polarizations of both the $\tau$ lepton and the hyperon $Y$, 
systematically addressing the theoretical uncertainty driven by the model dependence of form factors. 
Our analysis demonstrates that while this uncertainty leads to large variations in predicted rates, 
the polarization observables can potentially distinguish between the form factor models, 
and help to determine the chiral nature of the vectorial new physics interaction. 
Finally, we forecast the event rates for future facilities, revealing a strong dependence 
on the form factor models, 
with predicted yields ranging from potentially observable levels 
to rates far below current detection thresholds.

\end{abstract}


\maketitle

\section{Introduction}
\label{sec:intro}

The Standard Model (SM) of particle physics, despite its tremendous success, is known to be incomplete. Searches for new physics (NP) beyond the SM therefore constitute a major research thrust in particle physics.
A particularly compelling avenue is the search for charged lepton flavor violation (cLFV)~\cite{Calibbi:2017uvl,Ardu:2022sbt,Davidson:2022jai,Frau:2024rzt}. Since the rates for cLFV processes induced by neutrino mixing are suppressed to unobservable levels within the SM~\cite{Calibbi:2017uvl,Ardu:2022sbt}, any observation of a cLFV process would constitute a clear signal of NP.

In this work, we focus on the cLFV processes mediated by an $e-\tau$ transition coupled to a flavor-changing neutral current in the quark sector, specifically $e^-d \to\tau^- s$ transition. 
The effective interactions governing these process have been investigated through a variety of experimental probes. 
At high energies, these include the analyses of high-$p_T$ dilepton final states at the LHC~\cite{ATLAS:2018mrn,Angelescu:2020uug}, 
and potential future searches at the upcoming Electron-Ion Collider~\cite{Cirigliano:2021img}. In the low-energy regime, the primary probes are searches for cLFV tau decays, such as $\tau^- \to e^- K_S$, $\tau^- \to e^- K^{*0}$, and $\tau^{-} \to e^{-} \pi^{-} K^{+}$~\cite{Belle:2010rxj,Belle:2023ziz,Belle:2012unr,ParticleDataGroup:2024cfk}.
However, all these low-energy decay probes are exclusively \textit{mesonic}. The baryonic decays, 
which arise from the same underlying interaction, are kinematically inaccessible. 
To address this gap, we propose the quasi-elastic (QE) scattering process $e^- N \to \tau^- Y$ to probe the baryonic sector of these LFV interactions.

A complete determination of the underlying NP requires not only establishing its existence but also pinpointing the specific Dirac structures and Wilson coefficients (WCs) of the effective operators involved. However, the existing set of observables---including the cross sections of the new proposed scattering processes---is insufficient to fully determine all the parameters in the general effective Lagrangian as introduced in Eq.~\eqref{eq:Leff}. To provide the necessary new information, we investigate the final-state polarization of the produced $\tau$ lepton and hyperon $Y$. 
Though experimentally more demanding, these polarization observables are sensitive to different combinations of the effective operators and can therefore provide the new, independent constraints needed to fill the observational gap.

Connecting the observables considered in this work to the underlying NP---that is, determining the Dirac structures and their associated WCs---relies on a precise calculation of the hadronic matrix element for the $N \to Y$ transition. 
In the absence of direct calculations for these form factors, we consider the parametrizations derived from two distinct approaches. 
The first employs the SU(3) flavor symmetry to relate the $N \to Y$ transition form factors to the better-known $N \to N$ ones~\cite{Singh:2006xp,Sobczyk:2019uej}, 
neglecting the SU(3) breaking effects. The second utilizes theoretical calculations of the $Y \to N$ decay form factors, which must then be analytically continued from the timelike ($q^2>0$) to the spacelike ($q^2<0$) region required for scattering~\cite{Sobczyk:2019uej,Lai:2021sww,Lai:2022ekw,Kong:2023kkd,Yan:2024bce}. 
As we will demonstrate, this choice of model induces a strong dependence in all observables, altering the predicted scattering cross section by orders of magnitude and dramatically affecting the magnitudes and kinematic dependencies of the polarization observables.

This strong model dependence, while a challenge, also presents an opportunity for its resolution. In this work, we systematically investigate the final-state averaged polarization observables, examining their behavior across a range of NP scenarios and form factor parametrizations. 
We show that, under a specific NP assumption such as a purely vector interaction, 
certain combinations of hyperon polarization observables, \emph{e.g.}, $\langle P_{L,T}^\Sigma \rangle$ and $\langle P^\Lambda_L \rangle$, 
exhibit qualitatively different behaviors for the form factor models, separating them into distinct classes. Furthermore, we outline a strategy where 
the measurement of other polarization combinations, such as the tau polarization $\langle P_L^\tau \rangle$ and the hyperon polarization $\langle P_T^\Sigma \rangle$, can uniquely identify the chiral nature of the underlying vector operator, provided that the contributions from tensor operators are negligible.

The QE scattering processes we propose can be explored through a fixed-target experiment, thanks to the advanced capabilities of 
modern facilities such as the 12 GeV Continuous Electron Beam Accelerator Facility (CEBAF) at Jefferson Lab (JLab) and its potential upgrades~\cite{Arrington:2021alx}.
Using the current stringent constraints on the WCs from the cLFV tau decays as an input, 
we forecast the expected annual event rates for the QE scattering processes. 
Our results reveal a large spread in the predicted annual event rates, ranging from fewer than one to several hundred, with the variation driven mainly 
by the choice of form factor model. 

The remainder of this paper is organized as follows. In Sec.~\ref{sec:models}, we lay out our theoretical framework, detailing the effective Lagrangian, the construction of spin density matrices, and the various form factor parametrizations. In Sec.~\ref{sec:Numresults}, we present a comprehensive phenomenological analysis of the cross sections and polarization observables. We then assess the experimental prospects in Sec.~\ref{sec:Prospect}, using current constraints to project event rates. Finally, we summarize our results and offer our conclusions in Sec.~\ref{sec:conclusion}.

\section{Theoretical framework}
\label{sec:models}

\subsection{The low-energy effective Lagrangian}

The general low-energy effective Lagrangian involving $e^- d \to \tau^- s$ transitions can be written as
\begin{align}\label{eq:Leff}
	\mathcal{L}_{\text{eff}}=-\frac{4G_F}{\sqrt{2}}\sum_{\alpha}g_{\alpha}\mathcal{O}_{\alpha}+\text{h.c.}\,,
\end{align}
where $G_F$ is the Fermi constant, $\mathcal{O}_{\alpha}$ is the semi-leptonic operator listed in Table~\ref{tab:Operator}, and $g_{\alpha}$ is the corresponding effective WC. 
Often, the tensor operators  $\mathcal{O}^{RR}_{T}$ and $\mathcal{O}^{LL}_{T}$ are recast in the following form
\begin{align}
	\mathcal{O}^{LL,RR}_{T}=(\bar{\tau}\sigma_{\mu\nu}P_{L,R}e)(\bar{s}\sigma^{\mu\nu}d)\,,
\end{align}
making use of the identity $\sigma_{\mu\nu}\gamma_5=\frac{i}{2}\epsilon_{\mu\nu\rho\sigma}\sigma^{\rho\sigma}$. 
In contrast,  
the tensor operators $\mathcal{O}^{LR,RL}_{T}$ vanish identically as a consequence of Lorentz invariance.

\begin{table}[t]
	\renewcommand*{\arraystretch}{1.5}
	\tabcolsep=0.1cm
	\centering
	\caption{Operator $\mathcal{O}_{\alpha}$ of $\mathcal{L}_{\text{eff}}$ in Eq.~\eqref{eq:Leff}, where $P_{R,L}\!=\!(1\pm \gamma_5)/2$ denote the right- and left-handed projectors, and $\sigma^{\mu \nu}\!= \! i[\gamma^{\mu},\gamma^{\nu}]/2$ is the antisymmetric tensor. Note that coeff is short for coefficient.}
	\begin{tabular}{cccc}
		\hline \hline
		Coeff. &  Operator & Coeff. &  Operator  \\
		\hline		
		$g^{LL}_V$&$(\bar{\tau}\gamma_{\mu}P_{L}e)(\bar{s}\gamma^{\mu}P_{L}d)$ & $g^{RR}_V$&$(\bar{\tau}\gamma_{\mu}P_{R}e)(\bar{s}\gamma^{\mu}P_{R}d)$\\   
		$g^{LR}_V$&$(\bar{\tau}\gamma_{\mu}P_{L}e)(\bar{s}\gamma^{\mu}P_{R}d)$&
		$g^{RL}_V$&$(\bar{\tau}\gamma_{\mu}P_{R}e)(\bar{s}\gamma^{\mu}P_{L}d)$\\ 
		$g^{LL}_S$&$(\bar{\tau}P_{L}e)(\bar{s}P_{L}d)$&
		$g^{RR}_S$&$(\bar{\tau}P_{R}e)(\bar{s}P_{R}d)$\\   
		$g^{LR}_S$&$(\bar{\tau}P_{L}e)(\bar{s}P_{R}d)$&
		$g^{RL}_S$&$(\bar{\tau}P_{R}e)(\bar{s}P_{L}d)$\\ 
		$g^{LL}_T$&$(\bar{\tau}\sigma_{\mu\nu}P_{L}e)(\bar{s}\sigma^{\mu\nu} P_{L}d)$ &
		$g^{RR}_T$&$(\bar{\tau}\sigma_{\mu\nu}P_{R}e)(\bar{s}\sigma^{\mu\nu} P_{R}d)$\\                                                                                                  
		\hline \hline
	\end{tabular}	
		\label{tab:Operator}
\end{table}

\subsection{$\tau$- and $Y$-production density matrix}

Within the framework of the effective Lagrangian $\mathcal{L}_{\text{eff}}$, the scattering amplitude $\mathcal{M}$ for the QE 
processes $e^-+N\to \tau^-+Y$, with nucleons $N=n, p$ and hyperons $Y=\Lambda, \Sigma^0, \Sigma^+$, takes the general form
\begin{align}\label{eq:amplitude}
	\mathcal{M}=-\frac{4G_F}{\sqrt{2}}\sum_{a=S,V,T}\left(H^{L}_{a}L^{L}_{a}+H^{R}_{a}L^{R}_{a}\right)\,.
\end{align}
Here, the index $a$ runs over the scalar ($S$), vector ($V$), and tensor ($T$) interactions. The corresponding matrix elements of 
left- and right-chiral hadronic currents ($H^{L,R}_a$) and leptonic currents ($L^{L,R}_a$) are detailed in Appendix~\ref{app:matrix ele}.

From this amplitude, we construct the spin density matrix for the produced $\tau$ lepton. By summing over the unobserved helicities of the initial electron ($r$), initial nucleon ($s$), and final hyperon ($t$), it is given by~\cite{Haber:1994pe,Yan:2024bce}
\begin{align}\label{eq:rhoTau}
	\rho^\tau_{\lambda,\lambda^{\prime}}&=\sum_{r,s,t} \mathcal{M}(\lambda) \mathcal{M}^*(\lambda^\prime)\nonumber \\[0.02cm]
	&=8G^2_F\sum_{r,s,t}\sum_{a,a^\prime}\big\{H^{L}_{a}H^{L*}_{a^\prime}L^{L}_{a}(\lambda)L^{L*}_{a^\prime}(\lambda^\prime)\nonumber \\[0.02cm]
	&\qquad+H^{R}_{a}H^{R*}_{a^\prime}L^{R}_{a}(\lambda)L^{R*}_{a^\prime}(\lambda^\prime)\big\}\,,
\end{align} 
where $\lambda$ and $\lambda'$ are the helicities of the outgoing $\tau$. The second equation is obtained by noting that interference terms 
between the left- and right-chiral leptonic currents vanish (i.e., $L^{L}_{a}L^{R*}_{a^\prime}=0$) in the limit of a massless initial-state electron.

To explicit evaluate the $L^{L,R}_{a}(\lambda)L^{L,R*}_{a^\prime}(\lambda^\prime)$, we employ the Bouchiat-Michel formula~\cite{Bouchiat:1958yui,Michel:1959dvg,Haber:1994pe}, which projects the spin-dependent terms onto a spin-vector basis:
\begin{align}\label{eq:BMformulea}
	u_{\lambda^\prime}(k^\prime)\bar{u}_{\lambda}(k^\prime)=\frac{1}{2}\left[\delta_{\lambda\lambda^\prime}\!+\!\sum^3_{i=1}\gamma_5\slashed{s}_i\sigma^i_{\lambda\lambda^\prime}\right](\slashed{k}^\prime+m_\tau)\,.
\end{align}
In this expression, $k'$ is the four-momentum of the $\tau$ lepton, $\sigma^i$ are the Pauli matrices, and $s_i^{\mu}$ are three orthogonal spacelike four-vectors that describe the $\tau$ spin orientation.

Analogously, the spin density matrix for the produced hyperon $Y$ is defined by fixing its helicities $t, t'$ and summing over the remaining ones:
\begin{align}\label{eq:rhoY}
	\rho^Y_{t,t^{\prime}}&=\sum_{r,s,\lambda} \mathcal{M}(t) \mathcal{M}^*(t^\prime)\nonumber \\[0.02cm]
	&=8G^2_F\sum_{r,s,\lambda}\sum_{a,a^\prime}\big\{H^{L}_{a}(t)H^{L*}_{a^\prime}(t^\prime)L^{L}_{a}L^{L*}_{a^\prime}\nonumber \\[0.02cm]
	&\qquad+H^{R}_{a}(t)H^{R*}_{a^\prime}(t^\prime)L^{R}_{a}L^{R*}_{a^\prime}\big\}\,.
\end{align}
After parametrizing the hadronic currents in terms of form factors, $H^{L,R}_{a}(t)H^{L,R*}_{a^\prime}(t^\prime)$ can be expressed using a set of spin vectors $n_i^{\mu}$ for the outgoing hyperon $Y$, analogous to the leptonic case.

\subsection{Polarization vectors of the final lepton and hyperon}
\label{subsec:polarization}

The production density matrix $\rho$  
can be expanded in terms of the Pauli matrices $\sigma^i$ as
\begin{align}\label{eq:rhoP1}
	\rho_{\lambda\lambda^\prime}&=\delta_{\lambda\lambda^\prime} C+\sum_i\sigma^i_{\lambda\lambda^\prime}
	\Sigma_i \nonumber \\[0.02cm]
	&=C\left(\delta_{\lambda\lambda^\prime}+ \sum_i\sigma^i_{\lambda\lambda^\prime}P_i\right)\,,
\end{align}
where $P_i=\Sigma_i/C$ denote the three components of the polarization 
vector $\mathcal{P}^{\mu}\equiv \sum_i P_i s^{\mu}_i$~\cite{Kong:2023kkd}.  
The explicit expressions for $C$ and $\Sigma_i$ can be obtained by matching the density matrix $\rho$
in Eq.~\eqref{eq:rhoP1} to that in Eq.~\eqref{eq:rhoTau} for the lepton $\tau$ case, 
and to that in Eq.~\eqref{eq:rhoY} for the hyperon $Y$ case.

The differential cross section of the QE scattering with one polarized final-state particle 
is directly related to the density matrix $\rho_{\lambda\lambda^{\prime}}$, and takes the form~\cite{Haber:1994pe,Yan:2024bce} 
\begin{align}\label{eq:diff cross}
	d\sigma_{\lambda\lambda^\prime}=\frac{1}{4F}\rho_{\lambda\lambda^\prime}d\Phi(k,p;k^\prime, p^\prime)\,,
\end{align}
where $F=4 \sqrt{(p \cdot k)^2 - m_e^2 m_N^2}$ is the flux factor, 
$d\Phi$ is the phase space element, 
and the prefactor $1/4$ accounts for the spin average over the initial electron and nucleon. 

In the laboratory (Lab) frame, the polarized differential cross section can be expressed as  
\begin{align}\label{eq:sigmaP}
	d\sigma_{\lambda,\lambda^\prime}&=\frac{C(q^2)}{64\pi F E m_{N}}\left(\delta_{\lambda\lambda^\prime}
	+\sum_i\sigma^i_{\lambda\lambda^\prime}P_i(q^2)\right)dq^2 \nonumber \\[0.02cm]
	&=\frac{1}{2}\frac{d\sigma}{dq^2}\left(\delta_{\lambda\lambda^\prime}
	+\sum_i\sigma^i_{\lambda\lambda^\prime}P_i(q^2)\right)dq^2\,,
\end{align}
where $q=k-k^\prime$ is the momentum transfer, $E$ is the incident electron beam energy, 
and 
\begin{align}
	\frac{d\sigma}{dq^2}=\frac{C(q^2)}{32\pi F E m_{N}}
\end{align}
is the unpolarized differential cross section for the QE scattering 
in the Lab frame.

To explicitly evaluate the polarization components $P_i$ for the lepton $\tau$ and hyperon $Y$ in the Lab frame, 
we introduce the following two spin vectors $s_i$ and $n_i$.   
For the lepton $\tau$, its spin vectors are defined as
\begin{equation}\label{eq:basis_lepton}
	\begin{aligned}
		s_L^{\mu}&=\left(\frac{|\pmb{k}^\prime|}{m_{\tau}},\frac{k^{\prime0}\pmb{k}^\prime}{m_{\tau}|\pmb{k}^\prime|}\right), \\
		s_P^{\mu}&=\left(0,\frac{\pmb{k}^\prime\times \pmb{k}}{|\pmb{k}^\prime\times \pmb{k}|}\right), \\
		s_T^{\mu}&=\left(0, \frac{\pmb{k}^\prime\times(\pmb{k}\times \pmb{k}^\prime)}{|\pmb{k}^\prime\times(\pmb{k}\times \pmb{k}^\prime)|}\right),
	\end{aligned}
\end{equation}
and similarly, for the hyperon $Y$, we define
\begin{equation}\label{eq:basis_hadron}
	\begin{aligned}
		n_L^{\mu}&=\left(\frac{|\pmb{p}^\prime|}{m_{Y}},\frac{p^{\prime0}\pmb{p}^\prime}{m_{Y}|\pmb{p}^\prime|}\right), \\
		n_P^{\mu}&=\left(0, \frac{\pmb{p}^\prime\times \pmb{k}}{|\pmb{p}^\prime\times \pmb{k}|}\right), \\
		n_T^{\mu}&=\left(0, \frac{\pmb{p}^\prime\times(\pmb{k} \times \pmb{p}^\prime)}{|\pmb{p}^\prime\times(\pmb{k} \times \pmb{p}^\prime)|}\right),
	\end{aligned}
\end{equation}
indicating the longitudinal ($L$), perpendicular ($P$), and transverse ($T$) directions, respectively, 
defined with respect to their reaction planes.

Finally, to investigate the dependence of the polarization on the beam energy, 
we introduce the averaged polarization component along direction $i$ as~\cite{Graczyk:2004uy,Fatima:2020pvv}
\begin{align}
	\langle P^{\tau,Y}_i\rangle=\frac{\int_{q^2_{\text{min}}}^{q^2_{\text{max}}}P^{\tau,Y}_i(q^2)\frac{d\sigma}{dq^2}dq^2}{\int_{q^2_{\text{min}}}^{q^2_{\text{max}}}\frac{d\sigma}{dq^2}dq^2}\,.
\end{align} 

\subsection{Form factors}
\label{subsec:Cross section}

The hadronic matrix elements $H^{L,R}_{a}$ for the $N\to Y$ ($d\to s$ at quark level) transitions can be parameterized in terms of form factors.  
For the vector and axial-vector currents, their matrix elements are given by  
\begin{align}
	&\quad \langle Y(p^{\prime})|\bar{s}\gamma_{\mu}d|N(p)\rangle \nonumber \\
	&=\bar{u}(p^{\prime})\left[\gamma_\mu f_1(q^2)+i\sigma_{\mu\nu}\frac{q^\nu}{2m_Y}f_2(q^2)
	+\frac{q_\mu}{m_Y}f_3(q^2)\right]u(p)\label{eq:FormV}\,, \\
	&\quad	\langle Y(p^{\prime})|\bar{s}\gamma_{\mu}\gamma_5d|N(p)\rangle \nonumber \\
	&=\bar{u}(p^{\prime})\left[\gamma_\mu g_1(q^2)
	+i\sigma_{\mu\nu}\frac{q^\nu}{m_Y}g_2(q^2)+\frac{q_\mu}{m_Y}g_3(q^2)\right]\gamma_5u(p)\,, \label{eq:FormV5}
\end{align}
where $f_i$ and $g_i$ ($i=1, 2, 3$) denote the vector and axial-vector transition form factors, respectively.
An alternative definition of these matrix elements, employed in Ref.~\cite{Athar:2020kqn}, introduces a different set of form factors,
denoted by $\tilde{f}_i$ and $\tilde{g}_i$. The two conventions are related by
\begin{align}
	&f_1=\tilde{f}_1, \quad 
	&&f_2=\frac{2m_Y}{m_N+m_Y}\tilde{f}_2, \nonumber \\
	&f_3=\frac{2m_Y}{m_N+m_Y}\tilde{f}_3, \quad 
	&&g_1=\tilde{g}_1+2\frac{m_Y-m_N}{m_N+m_Y}\tilde{g}_2,\nonumber \\
	&g_2=-\frac{2m_Y}{m_N+m_Y}\tilde{g}_2, \quad 
	&&g_3=\frac{2m_Y}{m_N+m_Y}\tilde{g}_3\,.
\end{align}

From Eqs.~\eqref{eq:FormV} and \eqref{eq:FormV5}, the hadronic matrix elements of the scalar and pseudoscalar 
currents can be obtained via the equations of motion. As the form factors associated with the tensor current have, 
to the best of our knowledge, not yet been computed, we do not consider the corresponding hadronic matrix elements in this work.

Prior to parametrizing the form factors, we note that in charged‐current $N\to Y$ ($u\to s$) transitions 
the imposition of time‐reversal invariance renders all form factors $f_i$ and $g_i$ real, while 
G-parity, SU(3) flavor symmetry and the conserved vector current hypothesis require $f_3=g_2=0$~\cite{Fatima:2022tlf}.
In what follows, we adopt the identical set of symmetry assumptions for the neutral‐current $N\to Y$ ($d\to s$) 
transitions.

Under the assumption that the vector and axial-vector currents transform as members of the 
SU(3) flavor octet and that the baryon states $|N\rangle$ and $|Y\rangle$ likewise reside in the octet, 
each form factor can be decomposed into two functions, $D(q^2)$ and $F(q^2)$, weighted by the appropriate 
Clebsch-Gordan coefficients (see \textit{e.g.} Refs.~\cite{Athar:2020kqn, Cabibbo:1965zza} for details). 
In particular, one may write 
\begin{align}
	f_i(q^2)&=a F^V_i(q^2)+bD^V_i(q^2)\,, \\
	g_i(q^2)&=a F^A_i(q^2)+bD^A_i(q^2)\,, \quad i=1, 2, 3
\end{align}
where the coefficients $a$ and $b$ for each $N\to Y$ transition are listed in Table~\ref{tab:SU3}.

\begin{table}[t]
	\renewcommand*{\arraystretch}{1.5}
	\tabcolsep=1cm
	\centering
	\caption{Values of the coefficients $a$ and $b$.}
	\begin{tabular}{lcc}
		\hline \hline
		Transitions & $a$ & $b$ \\
		\hline		
		$n\to \Lambda $ & $-\frac{3}{\sqrt{6}}$ & $-\frac{1}{\sqrt{6}}$   \\
		$n\to \Sigma^0 $ & $\frac{1}{\sqrt{2}}$ & $-\frac{1}{\sqrt{2}}$    \\
		$p\to \Sigma^+ $ & $-1$ & 1    \\                                                                                 
		\hline \hline		                                                                                                
	\end{tabular}	
	\label{tab:SU3}
\end{table}

In the vector channel, the functions $D_{1,2}^V(q^2)$ and $F_{1,2}^V(q^2)$ are given by
\begin{align}
	F^V_{1,2}(q^2) &= f^p_{1,2}(q^2) + \frac{1}{2}f^n_{1,2}(q^2)\,, \label{eq:FV_def_refined} \\
	D^V_{1,2}(q^2) &= -\frac{3}{2}f^n_{1,2}(q^2)\,. \label{eq:DV_def_refined}
\end{align}
Here, $f^{p,n}_{1,2}(q^2)$ represent the Dirac ($i=1$) and Pauli ($i=2$) form factors for the proton and neutron.
Each $f_{1,2}^{p,n}(q^2)$ is, in turn, composed of the standard electromagnetic form factors, $F^{p,n}_{1,2}(q^2)$, and a contribution from the strange-quark vector current, $F^s_{1,2}(q^2)$~\cite{Athar:2020kqn,Ilma:2024lkp}:
\begin{align}
	f_{1,2}^{n,p}=\left(\frac{1}{2}-2\sin^2\theta_W\right)F_{1,2}^{n,p}-\frac{1}{2}F_{1,2}^{p,n}-\frac{1}{2}F_{1,2}^s\,,
\end{align}
where $\theta_W$ is the weak mixing angle. 

The electromagnetic pieces are then related to the Sachs' electric and magnetic form factors via
\begin{align}
	F_1^{p,n}(q^2)&=\frac{G_E^{p,n}(q^2)+\tau\,G_M^{p,n}(q^2)}{1+\tau}\,,\\
	F_2^{p,n}(q^2)&=\frac{G_M^{p,n}(q^2)-G_E^{p,n}(q^2)}{1+\tau}\,,
\end{align}
where $\tau=-q^2/(4m^2_N)$ and $m_N$ denotes the nucleon mass. 
For the form factors $G_{E,M}^{p,n}$, we compare three widely used parameterizations: 
those of Galster \emph{et al.}~\cite{Galster:1971kv}, the BBBA by Bradford \emph{et al.}~\cite{Bradford:2006yz}, 
and the BHLT by Borah \emph{et al.}~\cite{Borah:2020gte}. 
Details about these parameterizations are collected in Appendix~\ref{app:FFparam}.

The strange vector form factors $F^s_{1,2}$ can be written as~\cite{Garvey:1992cg}
\begin{align}
	F_1^s(q^2)&=\frac{1}{6}\langle r_s^2\rangle\,q^2\,F(q^2),\\
	F_2^s(q^2)&=\mu_s\,F(q^2)\,,
\end{align}
where $\langle r_s^2\rangle$ is the mean‐squared strange radius and $\mu_s$ the 
strange magnetic moment. 
The radius $\langle r_s^2\rangle$ is 
related to the strange electric charge radius \(\langle r_E^2 \rangle^s\) via
\begin{equation}
	\langle r_s^2 \rangle = \langle r_E^2 \rangle^s - \frac{3\mu_s}{2 m_N^2},
\end{equation}
and $F(q^2)$ is taken to be a modified dipole,
\begin{equation}
	F(q^2)
	=\left(1-\frac{q^2}{4m_N^2}\right)^{-1}
	\left(1-\frac{q^2}{m_v^2}\right)^{-2}\,,
\end{equation}
with the dipole mass $m_v$ fixed through
\begin{equation}
	m_v^2
	=12\,\mu_s\bigl(\langle r_M^2\rangle^s-\langle r_E^2\rangle^s\bigr)^{-1}\,.
\end{equation}
Here, $\langle r_M^2\rangle^s$ denotes the strange magnetic radius. 
Following the prescription of Ref.~\cite{Ilma:2024lkp}, 
we employ the lattice QCD results~\cite{Alexandrou:2019olr} for these parameters, namely
$\mu_s=-0.017$, $\langle r^2_M\rangle^s=-0.015$ $\rm{fm}^2$, 
and  $\langle r^2_E\rangle^s=0.0048$ $\rm{fm}^2$. 

\begin{table}[t!]
	\renewcommand*{\arraystretch}{1.5}
	\tabcolsep=1cm
	\centering
	\caption{ 
		Axial charges $g_{A}^{q}$ and axial masses $m_{Aq}$ used in the dipole parametrization 
		of $F_{A}^{q}$ (Eq.~\eqref{eq:FAq}), from Ref.~\cite{Alexandrou:2021wzv}.}
	\begin{tabular}{lcc}
		\hline \hline
		$q$ & $g^q_A$ & $m_{Aq}$ [GeV] \\
		\hline		
		$u $ & 0.859 & 1.187   \\
		$d$ & $-0.423$ & 1.168    \\
		$s$ & $-0.044$ & 0.992    \\                                                                                 
		\hline \hline		                                                                                                
	\end{tabular}	
	\label{tab:FAq}
\end{table}

In the axial-vector sector, the SU(3) functions $F^A_1$ and $D^A_1$ 
are similarly constructed from the nucleon axial-vector form factors, $g_1^p$ and $g_1^n$:
\begin{align}
	D^A_{1}(q^2) &=-\frac{3}{2}g^n_{1}(q^2)\,, \\
	F^A_{1}(q^2) &=g^p_{1}(q^2)+\frac{1}{2}g^n_{1}(q^2)\,.
\end{align} 
The nucleon axial form factors, in turn, admit a decomposition into isovector 
and strange components~\cite{Ilma:2024lkp},
\begin{align}
	g_{1}^{p,n}=\pm \frac{1}{2}F_A^{ iv}-\frac{1}{2}F_{A}^s\,,
\end{align}
with 
\begin{align}
	F_A^{ iv}(q^{2})&=F_A^u(q^{2})-F_A^d(q^{2})\,, \\
	F^q_A(q^{2})&=g_A^{q}\left(1-\frac{q^2}{m_{Aq}^2}\right)^{-2}\,.  \quad q=u,d,s \label{eq:FAq}
\end{align}
The axial charges $g_A^q$ and axial masses $m_{Aq}$ are taken from the lattice QCD results of Ref.~\cite{Alexandrou:2021wzv}, as summarized in Table~\ref{tab:FAq}. Although Ref.~\cite{Alexandrou:2021wzv} also provides 
a $z$-expansion fit for \(F_{A}^{q}(q^{2})\), for simplicity in the present analysis 
we retain only the dipole one.

For the form factor $g_3$, we adopt the relation given by Nambu~\cite{Nambu:1960xd},
\begin{align}\label{eq:g3_Nambu}
	g_3(q^2)=\frac{(m_N+m_Y)^2}{2(m_K^2-q^2)}g_1(q^2)\,,
\end{align}
where $m_K$ is the kaon mass.
This relation was originally formulated for charged-current $N\to Y$ transitions; 
we extend its application to the neutral-current channel by invoking isospin symmetry.
While other models exist, such as that of Marshak \emph{et al.}~\cite{Marshak:1969}, 
we focus on the Nambu relation for simplicity. A detailed comparison of the phenomenological impact of different $g_3$ models 
on the charged-current $N\to Y$ transitions can be found in Ref.~\cite{Fatima:2022tlf}.

Besides the approach outlined above, one may alternatively extract the hadronic matrix elements $H^{L,R}_{a}$ from their complex conjugates~\cite{Sobczyk:2019uej,Lai:2021sww}, which are themselves expressed in terms of the $Y\to N$
transition form factors. Because the kinematic domain probed in scattering processes ($q^2<0$) differs 
from that of semileptonic decays ($q^2>0$), a 
reliable theoretical treatment of scattering requires an analytic continuation of these form factors into the spacelike region.  
Accordingly, any parametrization employed must exhibit the requisite analyticity across both timelike and spacelike $q^2$ regions.
      
Guided by this requirement, we compare two parametrization schemes for the form factors in the $Y\to N$ transition. The first is based on leading-order chiral perturbation theory ($\chi$PT)~\cite{Tandean:2019tkm}, and the second follows the most recent QCD sum-rule (QCDSR) analysis~\cite{Zhang:2024ick}. While both approaches furnish complete results for the vector and axial-vector form factors, neither addresses the tensor-current contributions. Adopting our conventions for parametrizing the corresponding matrix elements, 
we find in the $\chi$PT scheme $f^\prime_2 = f^\prime_3 = g^\prime_2 = 0$,
whereas in the QCDSR framework $f^\prime_3 = g^\prime_2=g^\prime_3 = 0$. 
Here, $f^\prime_i(q^2)$ and $g^\prime_i(q^2)$ ($i$=1, 2, 3) denote the vector and axial-vector form factors in 
the $Y\to N$ transitions. 
For explicit expressions and further details of each model’s form factors, the reader is referred to Appendix~\ref{app:form factors}.   
   
 \section{Phenomenology analyses}
 \label{sec:Numresults}  
    
\subsection{Generic features of polarization observables}
\label{sec:obs}

The factors $C$ and $\Sigma_i$, which determine the polarization components 
$P_i$ (cf. Eq.~\eqref{eq:rhoP1}),  
can be expressed compactly as
\begin{align}
	C&=8G_F^2\sum g_{\alpha}g^*_{\beta}~\mathcal{A}_{\alpha-\beta}\,, \label{eq:dsigma} \\[0.02cm]
	\Sigma_i&=8G_F^2\sum g_{\alpha}g^*_{\beta}~\mathcal{A}^{\tau,Y}_{\alpha-\beta}\,. \label{eq:dsigma2}
\end{align} 
Here, $\mathcal{A}_{\alpha-\beta}$ denotes the reduced amplitude squared arising from the interference of 
operators $\mathcal{O}_{\alpha}$ and $\mathcal{O}_{\beta}$, 
while $\mathcal{A}^{\tau,Y}_{\alpha-\beta}$ encodes the corresponding polarization 
information of the outgoing \(\tau\) or $Y$. Since the explicit expressions for $\mathcal{A}_{\alpha-\beta}$ and $\mathcal{A}^{\tau,Y}_{\alpha-\beta}$ are lengthy and can be readily obtained using packages such as FeynCalc~\cite{Shtabovenko:2023idz} or Package-X~\cite{Patel:2016fam}, they are not detailed in this work.  
The sums over $\alpha,\beta$ span all the WCs in Table~\ref{tab:Operator}, 
since each operator may contribute to the scattering amplitude. 
However, as these WCs have not yet been determined, a fully general study of the kinematic dependence of
$P_{i}$ on $q^2$ and the beam energy $E$ is infeasible. 
Consequently, throughout this study, we adopt the simplifying assumption that only one operator $\mathcal{O}_\alpha$ in the effective Lagrangian of Eq.~\eqref{eq:Leff} is active at a time, and refer each such choice as a distinct NP scenario.

\begin{figure}[t]
	\centering
	\subfigure{\begin{minipage}{0.238\textwidth}
			\centering
			\includegraphics[width=1.64in]{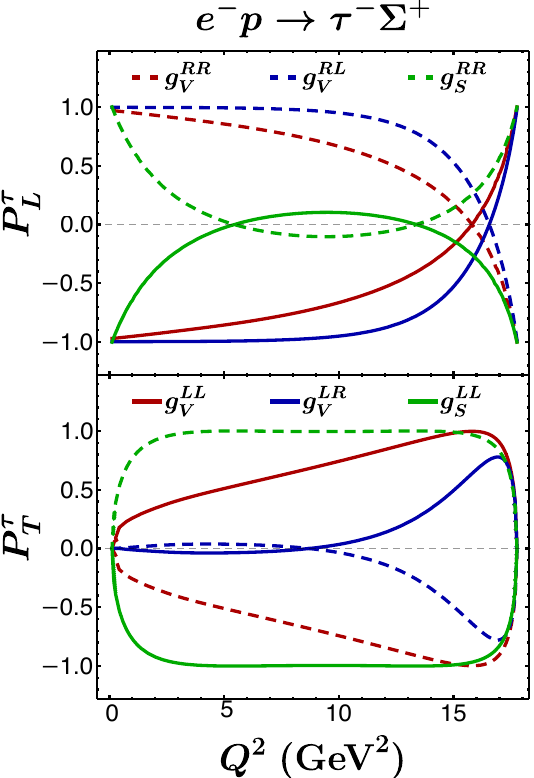}
	\end{minipage}}\hspace{0.018cm}
	\subfigure{\begin{minipage}{0.238\textwidth}
			\centering
			\includegraphics[width=1.65in]{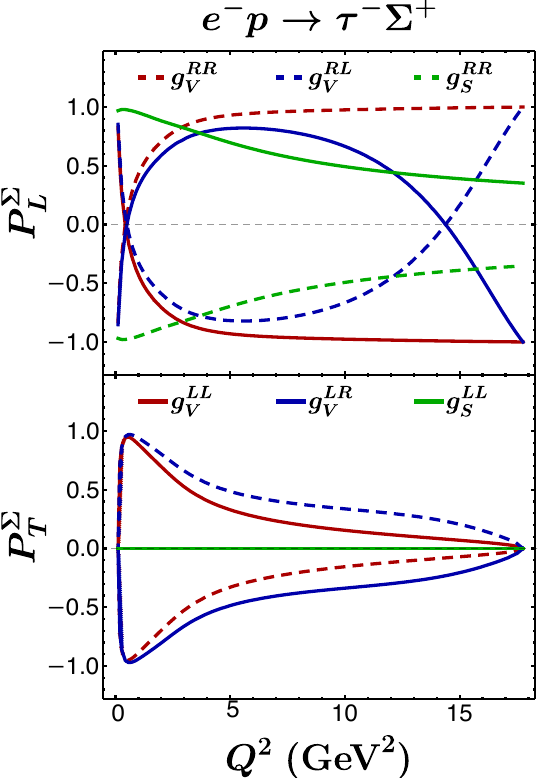}
	\end{minipage}}
	\caption{Variations of the polarization $P^{\tau,\Sigma}_L$ and $P^{\tau,\Sigma}_T$ with respect to $Q^2$ for various NP scenarios, 
		assuming the electron beam energy of $E=12$~GeV. For an illustration, we consider the process $e^-p\to \tau^-\Sigma^+$ 
		with the relevant electric and magnetic form factors taken from the BBBA parametrization~\cite{Bradford:2006yz}.} 
	\label{fig:NonMixCombine} 
\end{figure}

An examination of  $\mathcal{A}_{\alpha-\beta}$ and $\mathcal{A}^{\tau,Y}_{\alpha-\beta}$ 
reveals several generic features of the polarization $P_i$. 
These features hold for all the QE scattering processes considered herein and are independent of the specific 
parametrization schemes employed for the hadronic form factors.  
First, within the NP scenarios, the perpendicular polarization component $P^{\tau,Y}_P$ vanishes. 
This is due to the absence of terms involving the Levi-Civita tensor, $\varepsilon_{\{k\}\{k^\prime\}\{s_i\}\{p\}}$, 
a necessary condition for generating a non-zero $P_P$ in the Lab frame~\cite{Yan:2024bce}.
Second, the reduced amplitudes $\mathcal{A}_{\alpha\alpha}$ (where $\alpha=\beta$ due to the single operator assumption) are identical for NP scenarios corresponding to $g^{LL}_V$ and $g^{RR}_V$. In contrast, the associated polarization terms $\mathcal{A}^{\tau,Y}_{\alpha\alpha}$ differ by an overall sign. Consequently, the longitudinal ($P^{\tau,Y}_L$) and transverse ($P^{\tau,Y}_T$) polarization components also differ by a sign. An analogous relationship is also observed when comparing NP scenarios within the pairs ($g^{LR}_V$, $g^{RL}_V$), ($g^{LL}_S$, $g^{RR}_S$), and ($g^{LR}_S$, $g^{RL}_S$).  
In essence, simultaneously flipping the chiralities of both the lepton and quark currents of an operator $\mathcal{O}_{\alpha}$ 
results in an opposite sign for the induced $P_{L,T}$.
These sign relationships are dictated by the distinct chiral structures of the lepton and quark currents associated with these operators.

To illustrate the second feature concretely, we consider the QE scattering process 
$e^- p \to \tau^- \Sigma^+$, with the relevant hadronic form factors parameterized using the BBBA scheme for demonstration. 
Adopting a benchmark beam energy of $E=12$~GeV, we examine the $Q^2$ ($Q^2=-q^2$) dependence of the $P^{\tau,\Sigma}_L$ and $P^{\tau,\Sigma}_T$  for different NP scenarios. The results, depicted in Fig.~\ref{fig:NonMixCombine}, explicitly demonstrate the sign-flip behavior discussed above. As illustrated by comparing the solid and dashed curves of the same color within the figure, the signs of $P_{L,T}$ indeed reverse when the chiralities of both the lepton and quark currents are simultaneously flipped.

One may notice that the $P^{\tau,\Sigma}_L$ and $P^{\tau,\Sigma}_T$ induced by the scalar operators 
$\mathcal{O}^{LR}_S$ and $\mathcal{O}^{RL}_S$ 
are not explicitly depicted in Fig.~\ref{fig:NonMixCombine}. This absence is due to a third generic feature observed in these NP scenarios. Specifically, the $P^{\tau}_{L,T}$ induced by $\mathcal{O}^{LR}_S$ are identical to those induced by $\mathcal{O}^{LL}_S$. Conversely, and interestingly, the $P^{Y}_{L,T}$ induced by $\mathcal{O}^{LR}_S$ are equal to the negative of those induced by $\mathcal{O}^{LL}_S$. The polarization effects of $\mathcal{O}^{RL}_S$ can then be readily deduced from its established sign-flipped relationship with $\mathcal{O}^{LR}_S$ (cf. the second feature discussed above). 
Consequently, in light of the three established features, it is sufficient to restrict our explicit consideration 
to the NP scenarios corresponding to $g^{LL}_V$, $g^{LR}_V$, and $g^{LL}_S$.

A fourth, and final, feature pertains to the transverse polarization ($P^{Y}_T$) of the hyperon $Y$. This component is identically zero for all the NP scenarios involving the scalar operators. This null result arises because $P^{Y}_T$ is directly proportional to the kinematic factor $p\cdot n_T$, which vanishes in the Lab frame (cf. Eq.~\eqref{eq:basis_hadron}).

\begin{table}[tbp]
	\renewcommand*{\arraystretch}{1.5}
	\tabcolsep=0.34cm
	\centering
	\caption{Chosen values of the dimensionless scaling factor $\kappa$ for the $p\to \Sigma^+$ and $n\to \Lambda$ transitions, tabulated according to the NP scenario and the form factor parametrization (QCDSR and $\chi$PT); for the other three parametrizations used in our analysis, we set $\kappa=1$.}
	\begin{tabular}{cccccc}
		\hline \hline
		\multicolumn{3}{c}{$p\to \Sigma^+$} & \multicolumn{3}{c}{$n\to \Lambda$} \\
		\cmidrule(lr){1-3} \cmidrule(lr){4-6}
		NP & QCDSR & $\chi$PT & NP & QCDSR & $\chi$PT  \\
		\hline
		$g^{LL}_V$ & 10 & 2 & $g^{LL}_V$  & 20 & 5 \\
		$g^{LR}_V$ & 10 & 2 & $g^{LR}_V$  & 15 & 5 \\
		$g^{LL}_S$ & 1000 &10 & $g^{LL}_S$  & 2000 & 10 \\
		\hline \hline
	\end{tabular}
	\label{tab:Scaling}
\end{table}

\subsection{Differential cross section \\
	and $Q^2$-dependent polarization observables}
\label{subsec:Diffcross}

We now analyze the $Q^2$ evolution of the differential cross section and the polarization observables $P^{\tau,Y}_i$, extending beyond their generic behaviors discussed above. In particular, we analyze the sensitivity of these observables to various NP scenarios and to the chosen parametrization of the form factors. Given the direct relationship between the $n\to \Sigma^0$ and $p \to \Sigma^+$ form factors under the SU(3) symmetry (see Table~\ref{tab:SU3}), we restrict our subsequent analysis to the $p \to \Sigma^+$ and $n\to \Lambda$ transitions.

\begin{figure}[t]
	\centering
	\subfigure{\begin{minipage}{0.238\textwidth}
			\centering
			\includegraphics[width=1.64in]{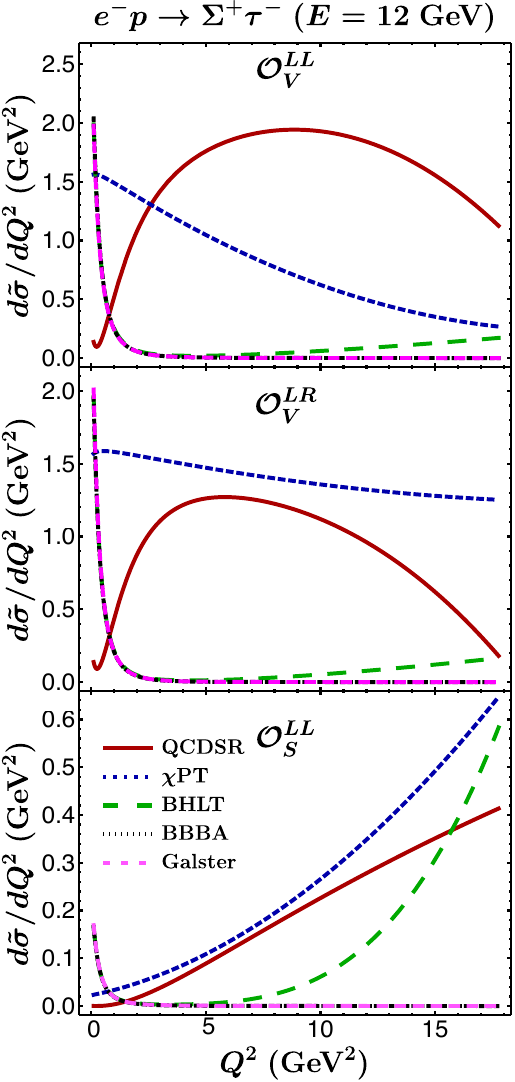}
	\end{minipage}}\hspace{0.018cm}
	\subfigure{\begin{minipage}{0.238\textwidth}
			\centering
			\includegraphics[width=1.64in]{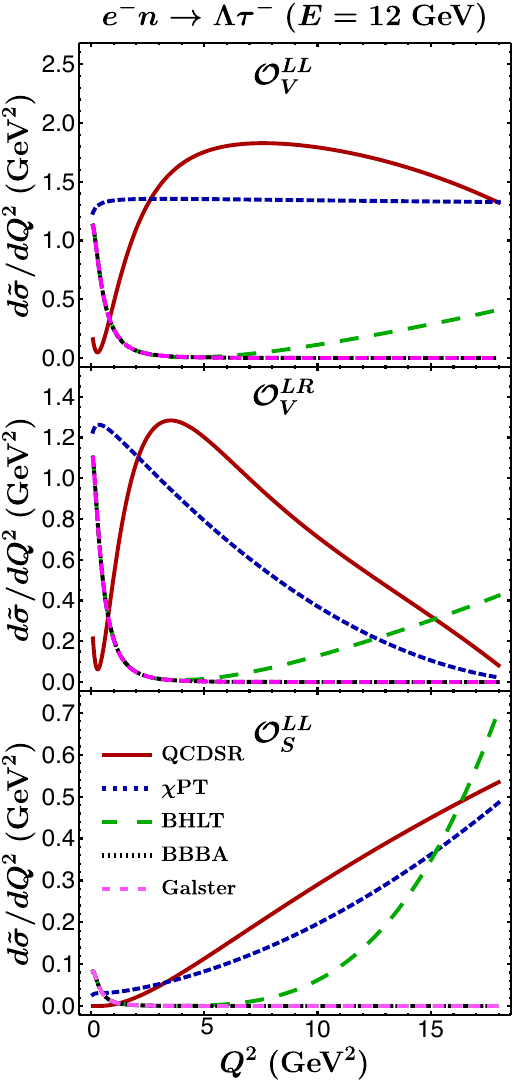}
	\end{minipage}}
	\caption{The reduced differential cross section $\tilde{\sigma}$ as a function of $Q^2$ for various NP scenarios and form factor parametrizations. The reduced cross section is defined by the relation $\sigma = \kappa G^2_F |g_{\alpha}|^2 \tilde{\sigma}/(16\pi m_N^2)$, where $\sigma$ is the full cross section and $\kappa$ is a dimensionless scaling factor. } 
	\label{fig:DiffCross} 
\end{figure}

\begin{figure*}[!hpt]
	\centering
	\includegraphics[width=0.96\textwidth]{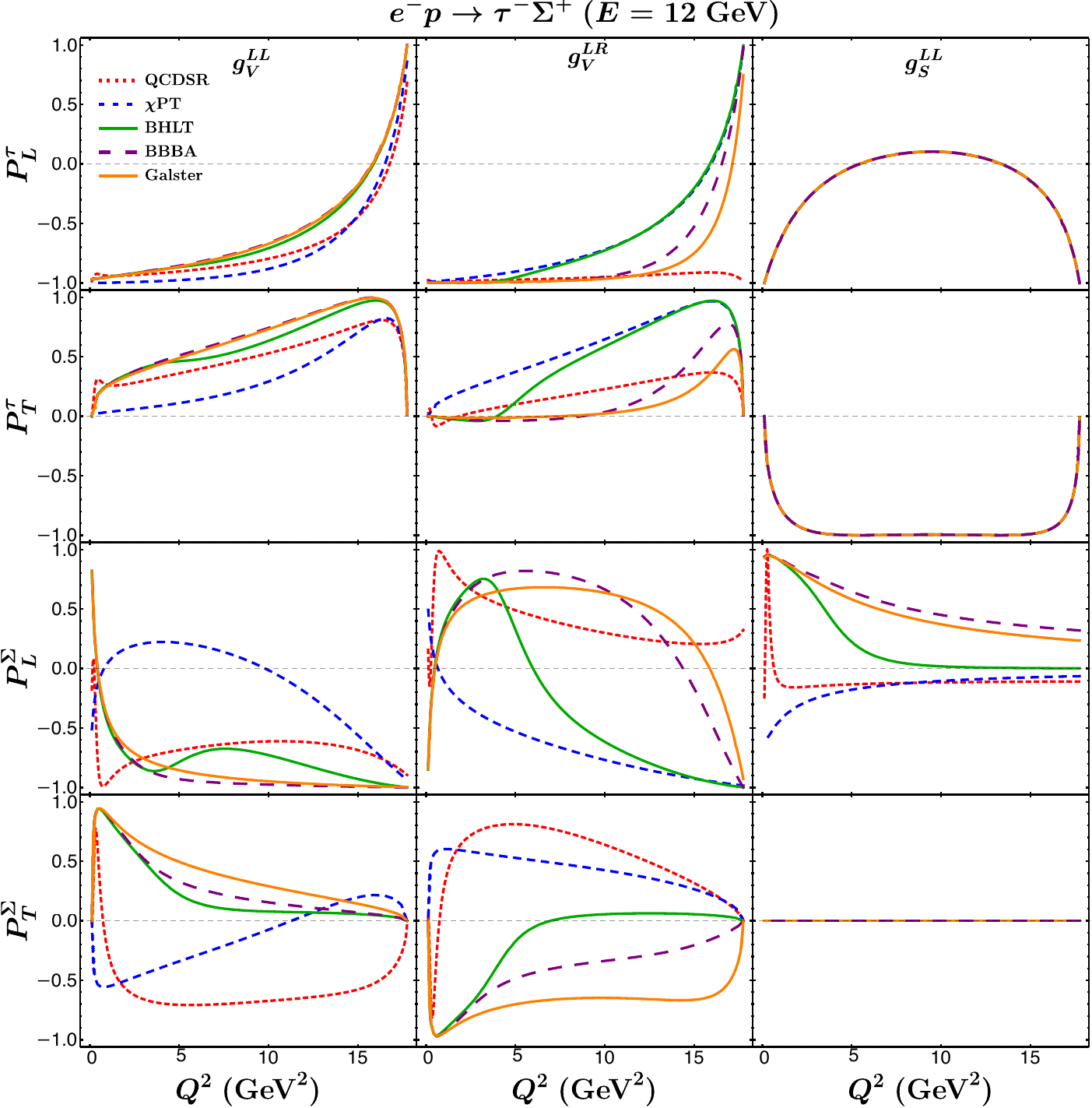}
	\caption{The $Q^2$ dependence of the $P_{L,T}$ polarization observables for the final-state $\tau^-$ and $\Sigma^+$ in the scattering process $e^- p \to \tau^- \Sigma^+$, assuming a beam energy of $E=12$~GeV. The plots compare predictions from different NP scenarios (columns) and form factor parametrizations (line styles).} 
	\label{fig:PSig} 
\end{figure*} 
 
\begin{figure*}[!hpt]
	\centering
	\includegraphics[width=0.96\textwidth]{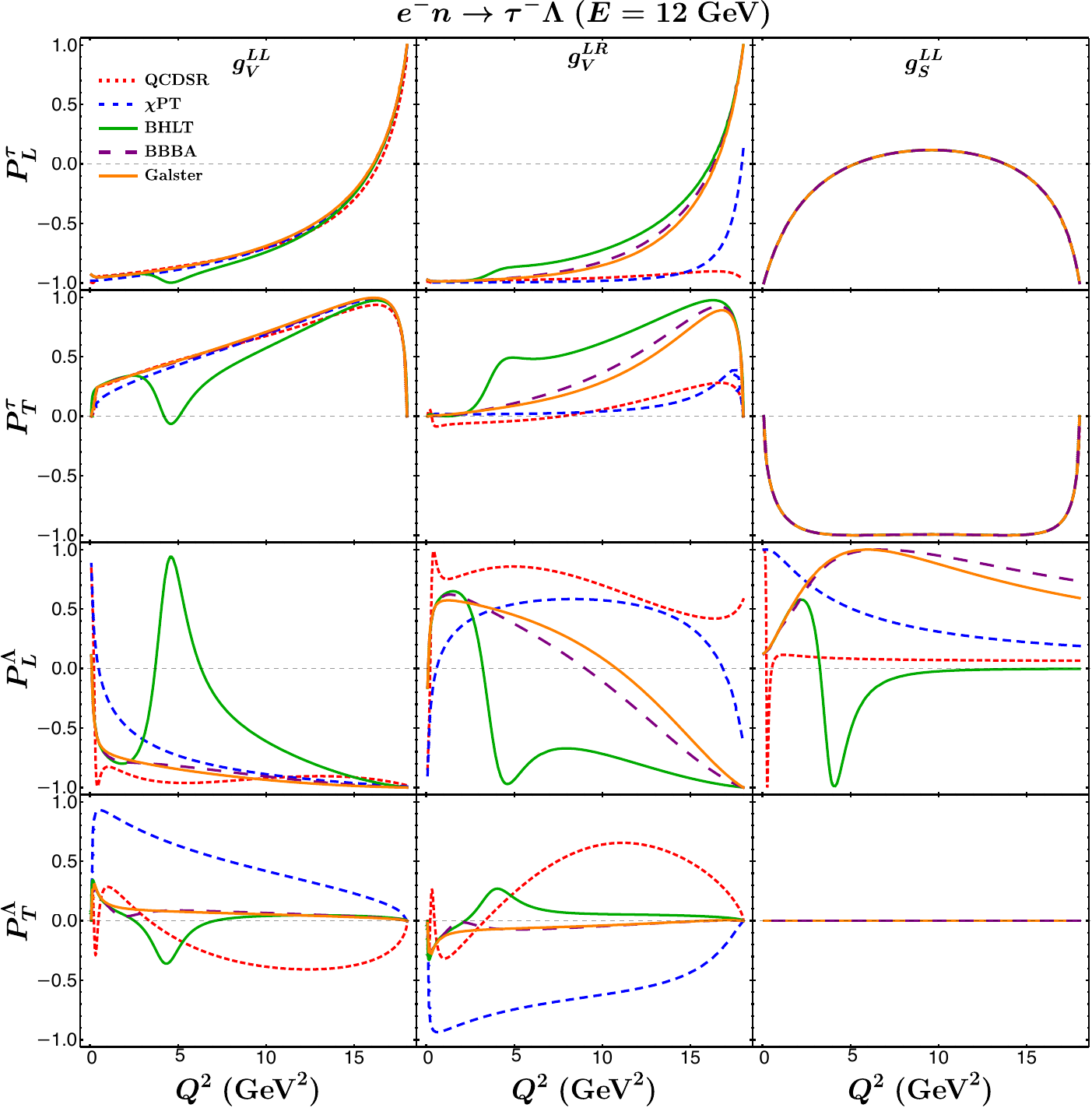}
	\caption{The $Q^2$ dependence of the $P_{L,T}$ polarization observables for the final-state $\tau^-$ and $\Lambda$ in the scattering process $e^- n \to \tau^- \Lambda$, assuming a beam energy of $E=12$~GeV. The plots compare predictions from different NP scenarios (columns) and form factor parametrizations (line styles).} 
	\label{fig:PLam} 
\end{figure*}  

We begin our analysis with the differential cross section, shown in Fig.~\ref{fig:DiffCross}, from which several key trends emerge. The BBBA and Galster parametrizations show remarkable consistency with each other across all NP scenarios and transition processes. In comparison, the BHLT scheme aligns with them at low $Q^2$ ($\lesssim 5~\text{GeV}^2$) but deviates at higher $Q^2$, with the disparity becoming most significant for the $g_S^{LL}$ scenario. By far, the largest cross sections are predicted by the QCDSR and $\chi$PT schemes; it must be emphasized that this visual dominance understates their true scale, as the values shown have already been suppressed by the scaling factors $\kappa$ listed in Table~\ref{tab:Scaling}. 
Finally, assuming a universal coupling strength (\emph{e.g.}, $g_\alpha = 1$), the contribution from the scalar $g_S^{LL}$ operator is dominant for the QCDSR, $\chi$PT, and BHLT models, whereas the BBBA and Galster predictions show little variation between the different NP scenarios.
 
The $Q^2$ dependence of the final-state $\tau$ and hyperon polarizations is presented in Fig.~\ref{fig:PSig} for the $e^- p \to \tau^- \Sigma^+$ process and in Fig.~\ref{fig:PLam} for $e^-n\to \tau^-\Lambda$. Focusing on the $P^{\tau}_{L,T}$ for the $p \to \Sigma^+$ transition (top two rows of Fig.~\ref{fig:PSig}), we find the behavior is highly dependent on the NP scenario. In the $g_V^{LL}$ case, all form factor models yield qualitatively similar predictions. For the $g_S^{LL}$ scenario, the predictions are entirely independent of the form factors, resulting in identical curves for $P^{\tau}_{L,T}$. In contrast, in the $g_V^{LR}$ scenario, the choice of form factors leads to qualitatively different predictions. 
For the longitudinal polarization $P^{\tau}_{L}$, the QCDSR model yields a nearly constant $P^{\tau}_{L}\approx -1$ across the entire $Q^2$ range, whereas all other models predict a rise towards positive values. For the transverse polarization $P^{\tau}_{T}$, the $\chi$PT and BHLT models provide similar results that are systematically larger than the predictions from the other three models.

Turning to the hyperon polarizations, $P^{\Sigma}_{L,T}$, shown in the bottom two rows of Fig.~\ref{fig:PSig}, 
we find a particularly strong sensitivity to the form factor models.
Take the longitudinal polarization, $P^{\Sigma}_L$. 
In the low-$Q^2$ region ($Q^2 \lesssim 1\,\text{GeV}^2$), for instance, the predictions from the QCDSR and $\chi$PT models
are qualitatively different from those of the BHLT, BBBA, and Galster models. 
This feature holds across all NP scenarios, with the QCDSR prediction, in particular, exhibiting multiple zero-crossings near $Q^2=0$. 
The transverse polarization, $P^{\Sigma}_T$, exhibits a similarly strong dependence on the form factor models.
At very low $Q^2$ and in the presence of vector operators (left and middle columns), the $\chi$PT prediction for $P^{\Sigma}_T$ has an opposite sign compared to the other models.
As $Q^2$ increases towards $1\,\text{GeV}^2$, the QCDSR prediction then becomes the distinct one, 
separating from the cluster of BHLT, BBBA, and Galster models.
At still higher $Q^2$, these latter three models yield distinguishable predictions for $P^{\Sigma}_T$, too. 
Interestingly, this distinction between the BBBA and Galster predictions is absent in the corresponding cross sections, 
where their results are nearly identical (see Fig.~\ref{fig:DiffCross}).

The $P^{\tau}_{L,T}$ for the $n \to \Lambda$ transition (top two rows of Fig.~\ref{fig:PLam}) share general features with the $p \to \Sigma^+$ case but also exhibit distinct model dependencies.
A distinguishing feature of the BHLT model, for instance, is the presence of localized structures near $Q^2 \approx 4.5\,\text{GeV}^2$.
In the $g_V^{LL}$ scenario, this model predicts a prominent dip in $P_T^\tau$ and a smaller one in $P_L^\tau$, while a corresponding bump appears for both observables in the $g_V^{LR}$ scenario.
Moreover, in the $g_V^{LR}$ scenario, the predictions for $P^{\tau}_T$ from the $\chi$PT and QCDSR 
models are nearly degenerate and lie systematically below those of the BHLT, BBBA, and Galster models.
     
The $P^{\Lambda}_{L,T}$ for the $n \to \Lambda$ transition (bottom two rows of Fig.~\ref{fig:PLam}) 
also exhibit a complex sensitivity to the form factor models.
As an example, the QCDSR prediction for the $P^{\Lambda}_{T}$ features three zero-crossings in the NP scenario 
involving the vector operators. Another notable feature is the appearance of localized structures, particularly in the BHLT model. 
For the $P^{\Lambda}_{L}$, these manifest as sharp structures near $Q^2 \approx 2\,\text{GeV}^2$ and $Q^2 \approx 4.5\,\text{GeV}^2$.
For the$P^{\Lambda}_{T}$, corresponding features appear near $Q^2 \approx 0\,\text{GeV}^2$ and $Q^2 \approx 4.5\,\text{GeV}^2$.
Finally, in contrast to the $p \to \Sigma^+$ transition, the distinction between the BBBA and Galster predictions is less pronounced here.
This distinction is most apparent for $P^{\Lambda}_{L}$ only, where a small separation between the BBBA and Galster predictions is visible at larger $Q^2$ in the $g_V^{LR}$ and $g_S^{LL}$ scenarios.
 
 \begin{figure}[tp]
 	\centering
 	\subfigure{\begin{minipage}{0.238\textwidth}
 			\centering
 			\includegraphics[width=1.68in]{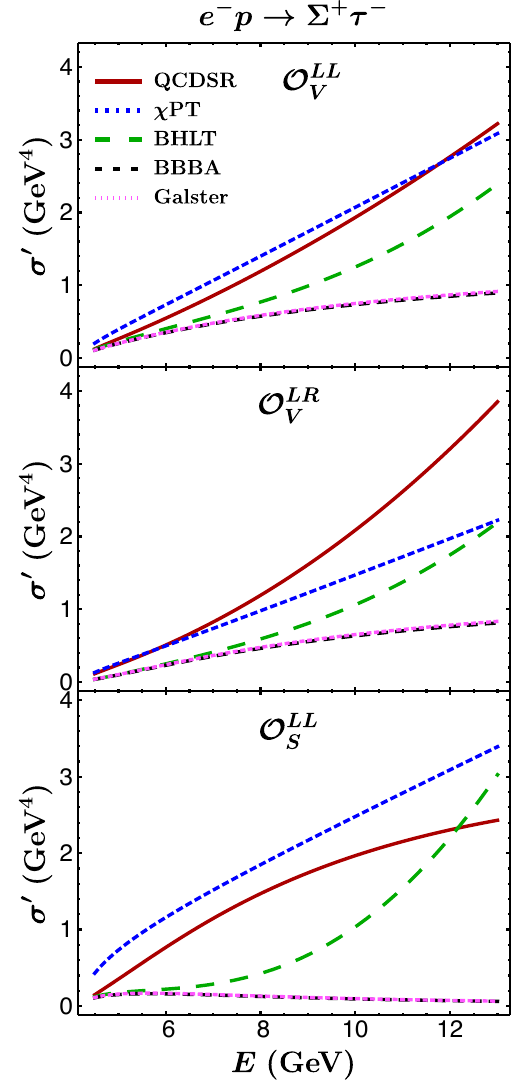}
 	\end{minipage}}\hspace{0.018cm}
 	\subfigure{\begin{minipage}{0.238\textwidth}
 			\centering
 			\includegraphics[width=1.62in]{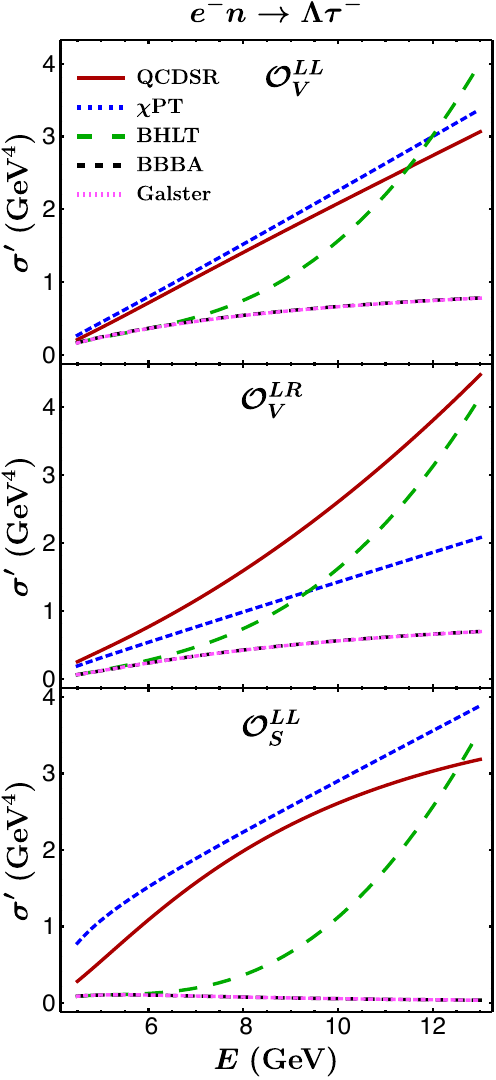}
 	\end{minipage}}
 	\caption{The reduced total cross section $\sigma^\prime$ as a function of the beam energy $E$ for various NP scenarios and form factor parametrizations. The reduced cross section is defined by the relation $\sigma = \kappa^\prime G^2_F |g_{\alpha}|^2 \sigma^\prime/(16\pi m_N^2)$, where $\sigma$ is the full cross section and $\kappa^\prime$ is a dimensionless scaling factor.} 
 	\label{fig:TotalCross} 
 \end{figure} 
 
 \begin{table}[!tbhp]
 	\renewcommand*{\arraystretch}{1.4}
 	\tabcolsep=0.34cm
 	\centering
 	\caption{Chosen values of the dimensionless scaling factor $\kappa^\prime$ for the $p\to \Sigma^+$ and $n\to \Lambda$ transitions, tabulated according to the NP scenario and the form factor parametrization (QCDSR and $\chi$PT); for the other three parametrizations used in our analysis, we set $\kappa^\prime=1$.}
 	\begin{tabular}{cccccc}
 		\hline \hline
 		\multicolumn{3}{c}{$p\to \Sigma^+$} & \multicolumn{3}{c}{$n\to \Lambda$} \\
 		\cmidrule(lr){1-3} \cmidrule(lr){4-6}
 		NP & QCDSR & $\chi$PT & NP & QCDSR & $\chi$PT  \\
 		\hline
 		$g^{LL}_V$ & 100 & 10 & $g^{LL}_V$  & 200 & 40 \\
 		$g^{LR}_V$ & 50 & 25 & $g^{LR}_V$  & 50 & 25 \\
 		$g^{LL}_S$ & 1500 &15 & $g^{LL}_S$  & 3000 & 10 \\
 		\hline \hline
 	\end{tabular}
 	\label{tab:Scaling2}
 \end{table}
 
\subsection{Total cross section and average polarizations}
\label{subsec:totalcross}
  
Having established the results at the benchmark energy in the previous subsection, we now investigate the dependence of the total cross section $\sigma$ and the average final-state polarizations $\langle P_i \rangle$ on the beam energy $E$. 

We begin with the total cross section, whose energy dependence is presented in Fig.~\ref{fig:TotalCross} for the processes $e^-p\to \Sigma^+\tau^-$ (left column) and $e^-n\to \Lambda\tau^-$ (right column). The results reveal two distinct behaviors depending on the form factor parametrization. The QCDSR, $\chi$PT, and BHLT models all predict a cross section that grows steeply with energy $E$ for all operators and processes considered. In contrast, the BBBA and Galster models yield only a mild rise for vector operators ($\mathcal{O}_V^{LL,LR}$) and a negligible, nearly flat cross section for the scalar operator ($\mathcal{O}_S^{LL}$). 
It should be noted that, similar to the differential cross section, the values shown for the total cross section have been scaled by the factors $\kappa^\prime$ listed in Table~\ref{tab:Scaling2}.

Besides the energy dependence, a comparison between the two processes reveals further model-dependent features. For the QCDSR, $\chi$PT, and BHLT models, the cross section for $e^-n\to \Lambda\tau^-$ is generally larger than or comparable to that for $e^-p\to \Sigma^+\tau^-$. In addition, 
the BBBA and Galster parametrizations are simpler in this regard, consistently yielding comparable cross sections for both processes in all NP scenarios.
  
\begin{figure*}[!hpt]
	\centering
	\includegraphics[width=0.96\textwidth]{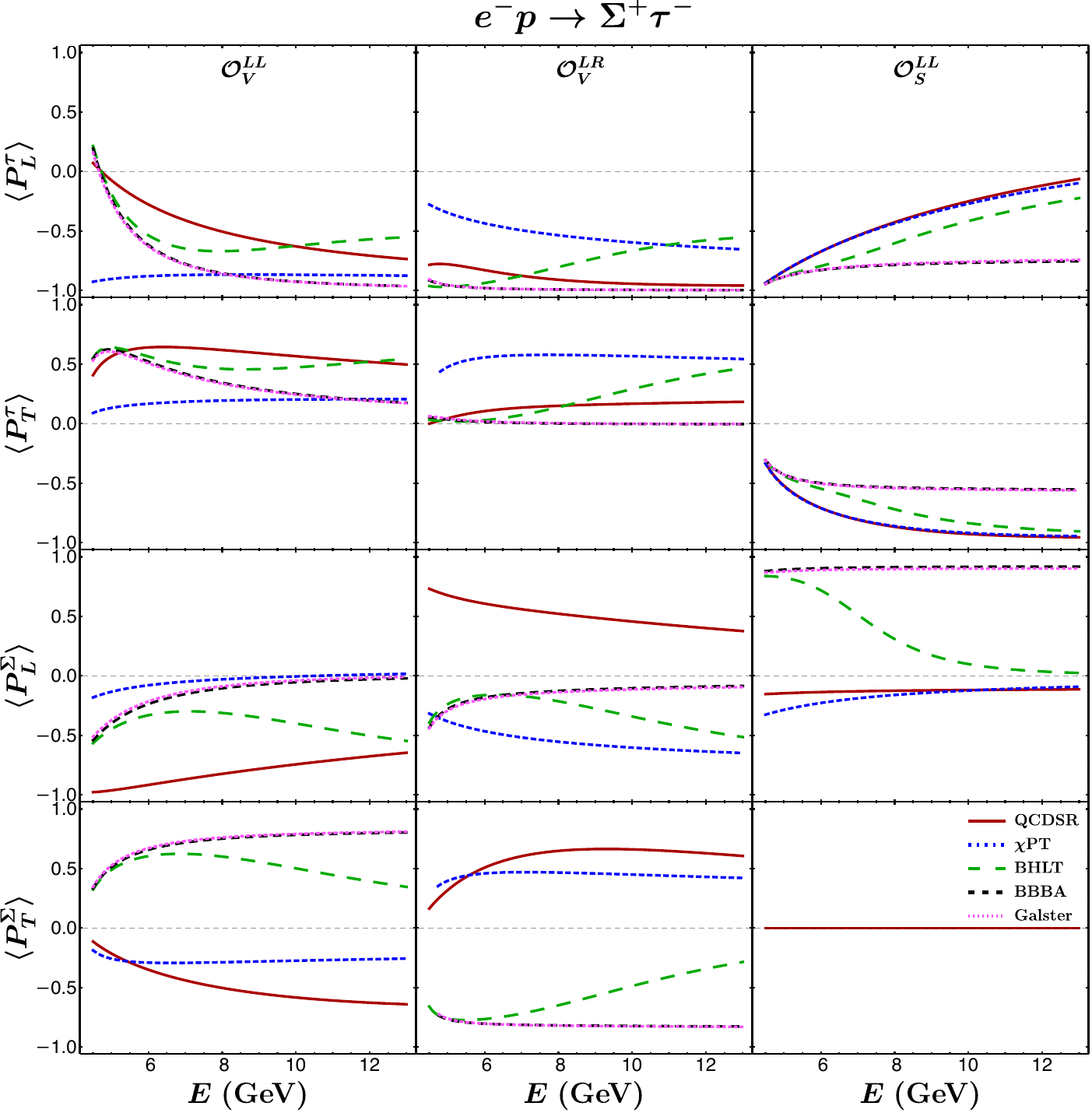}
	\caption{The dependence of the average polarization observables $\langle P_{L,T}\rangle$ on the incoming electron beam energy, $E$, for the process $e^- p \to \Sigma^+ \tau^-$. The plots compare predictions from different NP scenarios (columns) and form factor parametrizations (line styles).} 
	\label{fig:AvgPSig} 
\end{figure*}  
 
\begin{figure*}[!hpt]
	\centering
	\includegraphics[width=0.96\textwidth]{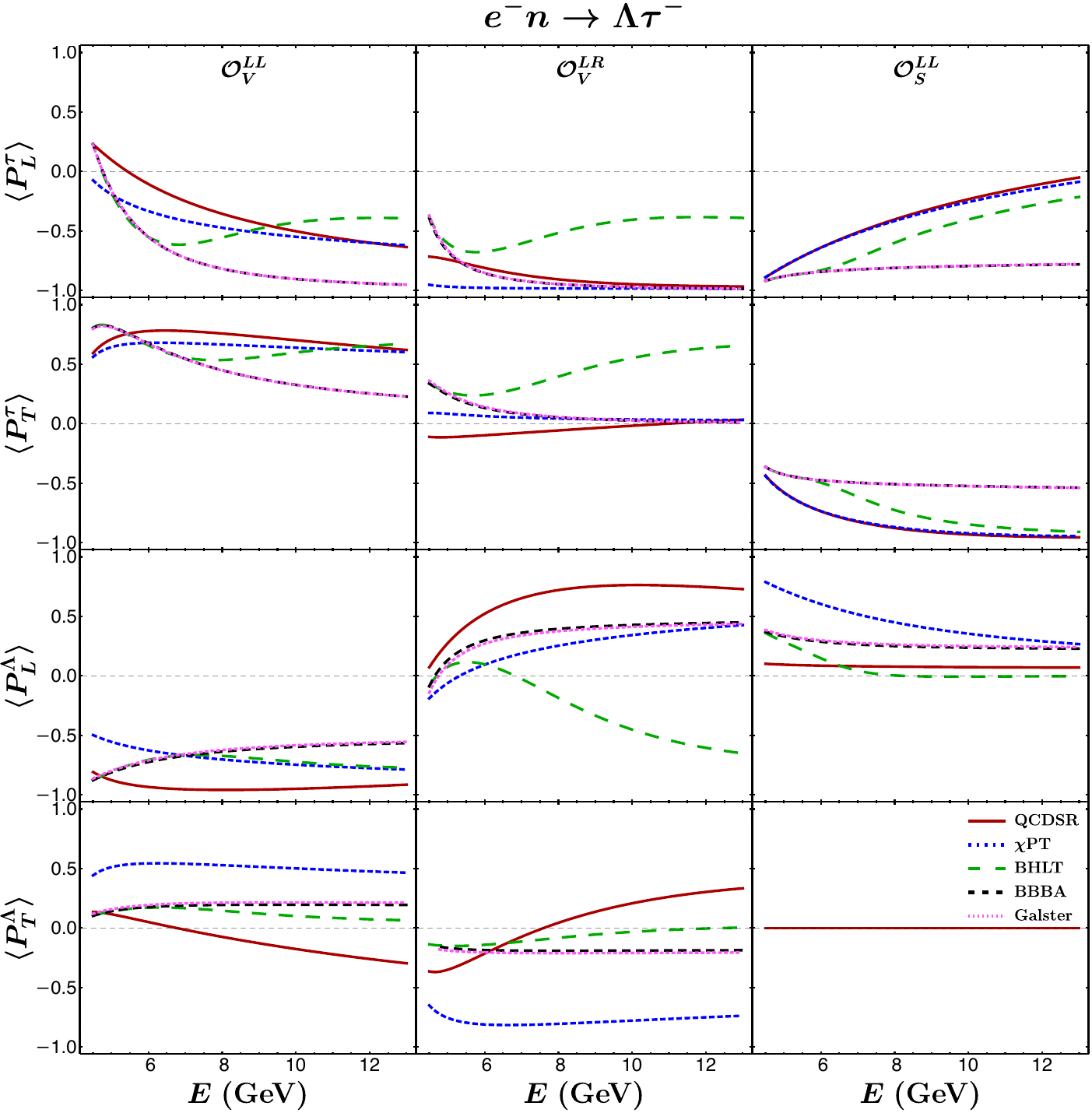}
	\caption{The dependence of the average polarization observables $\langle P_{L,T}\rangle$ on the incoming electron beam energy, $E$, for the process $e^- n \to \Lambda \tau^-$. The plots compare predictions from different NP scenarios (columns) and form factor parametrizations (line styles).} 
	\label{fig:AvgPLam} 
\end{figure*}  

We now turn to the averaged polarizations $\langle P_{L,T}\rangle$, whose energy dependence is presented in Fig.~\ref{fig:AvgPSig} for the process $e^-p\to \Sigma^+\tau^-$ and in Fig.~\ref{fig:AvgPLam} for $e^-n\to \Lambda\tau^-$. Beyond the strong overall model dependence, a combined analysis of the polarizations, $\langle P_{L,T}^\Sigma \rangle$ and $\langle P_{L,T}^\Lambda \rangle$, offers probable 
strategies for form factor model discrimination.

For the vectorial operator $\mathcal{O}_V^{LR}$, a measurement of the $\Sigma^+$ polarizations alone separates the models into three distinct classes. Specifically, QCDSR predicts positive values for both $\langle P_L^\Sigma \rangle$ and $\langle P_T^\Sigma \rangle$, whereas $\chi$PT predicts a negative $\langle P_L^\Sigma \rangle$ and a positive $\langle P_T^\Sigma \rangle$. In contrast, the remaining models (BBBA, Galster, and BHLT) predict both polarizations to be negative. 
This third group of models can then be resolved by a measurement of $\langle P_L^\Lambda \rangle$, as the BHLT model predicts a negative value, whereas the BBBA and Galster models predict positive values.
For the scalar operator $\mathcal{O}_S^{LL}$, the models also separate into three classes based on the behavior of $\langle P_L^\Sigma \rangle$. The QCDSR and $\chi$PT models predict negligible polarizations, $\langle P_{L,T}^{\Sigma} \rangle \approx -0.1$. The BBBA and Galster models predict a large and nearly constant longitudinal polarization, $\langle P_L^\Sigma \rangle \approx 1$. The BHLT model is uniquely identified by predicting a large but energy-dependent $\langle P_L^\Sigma \rangle$ that decreases as $E$ increases. Finally, for the vector operator $\mathcal{O}_V^{LL}$, the sign of the transverse polarization $\langle P_T^\Sigma \rangle$ provides a clean separation: it is negative for QCDSR and $\chi$PT, but positive for the other three models.

Beyond discriminating between form factor models, a multi-step procedure can be used to identify the specific NP operator. The first step distinguishes scalar from vector operators. As established in the analysis of Fig.~\ref{fig:AvgPSig}, the scalar $\mathcal{O}_S^{LL}$ operator yields $\langle P_T^\Sigma \rangle=0$, 
whereas all vector scenarios in Table~\ref{tab:sign_combinations} produce a non-zero transverse polarization. Thus, a measurement of a non-zero $\langle P_T^\Sigma \rangle$ would strongly indicate a vector interaction. 
Once an interaction has been identified as vectorial, a second step can be taken to determine its specific type. As detailed in Table~\ref{tab:sign_combinations}, this is achieved by measuring the sign of an additional observable, $\langle P_L^\tau \rangle$. The method relies on the one-to-one mapping between the sign combination of $(\langle P_L^\tau \rangle, \langle P_T^\Sigma \rangle)$ and the four vector operators within each form factor model's framework. 
For instance, assuming the $\chi$PT model, a measurement of these signs would uniquely determine whether the interaction is of the type $\mathcal{O}_V^{LL}$, $\mathcal{O}_V^{LR}$, $\mathcal{O}_V^{RR}$, or $\mathcal{O}_V^{RL}$. 
Therefore, if a specific model is taken as a baseline, a sequence of polarization measurements allows for an unambiguous determination of the underlying vector NP interaction, provided that contributions from the tensor operators are also negligible.
 
\begin{table}[!tp]
	\renewcommand*{\arraystretch}{1.3}
	\tabcolsep=0.43cm
	\centering
	\caption{Signatures from the polarization observables $\langle P_L^\tau \rangle$ and $\langle P_T^\Sigma \rangle$ used to identify vector NP operators in the process $e^-p\to \Sigma^+\tau^-$ with a proper (large) $E$. The predictions shown are representative of two distinct model families: \{QCDSR, $\chi$PT\} and \{BBBA, BHLT, Galster\}.}
	\label{tab:sign_combinations}
	\begin{tabular}{lccc}
		\hline \hline
		Model & NP  & Sign of $\langle P_L^\tau \rangle$ & Sign of $\langle P_T^\Sigma \rangle$ \\
		\midrule
		& $g_V^{LL}$ & $-$ & $-$  \\
		\cmidrule(lr){2-4}
		QCDSR &  $g_V^{LR}$                  & $-$& $+$ \\
		\cmidrule(lr){2-4}
		$\chi$PT &  $g_V^{RR}$                  & $+$& $+$ \\
		\cmidrule(lr){2-4}
		&  $g_V^{RL}$                  & $+$& $-$ \\
		\midrule
		BBBA  & $g_V^{LL}$ & $-$ & $+$  \\
		\cmidrule(lr){2-4}
		BHLT&  $g_V^{LR}$                  & $-$& $-$ \\
		\cmidrule(lr){2-4}
		Galster &  $g_V^{RR}$                  & $+$& $-$ \\
		\cmidrule(lr){2-4}
		&  $g_V^{RL}$                  & $+$& $+$ \\
		\botrule
	\end{tabular}
\end{table}

\begin{figure}[!b]
	\centering
	\subfigure{\begin{minipage}{0.238\textwidth}
			\centering
			\includegraphics[width=1.68in]{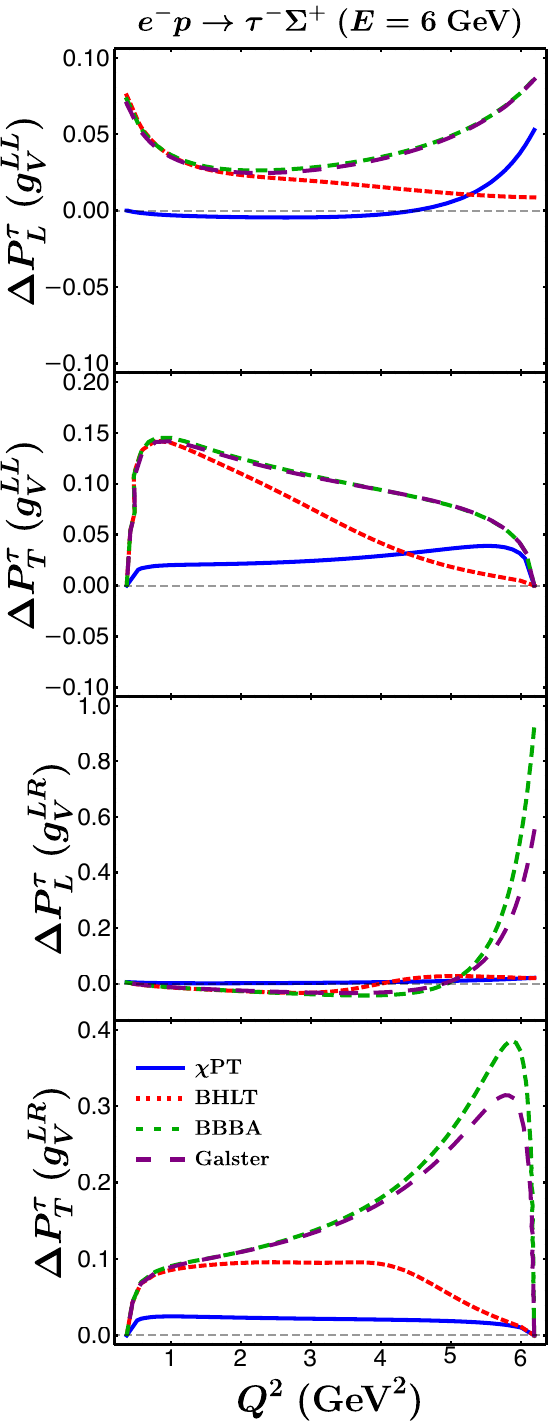}
	\end{minipage}}\hspace{0.018cm}
	\subfigure{\begin{minipage}{0.238\textwidth}
			\includegraphics[width=1.64in]{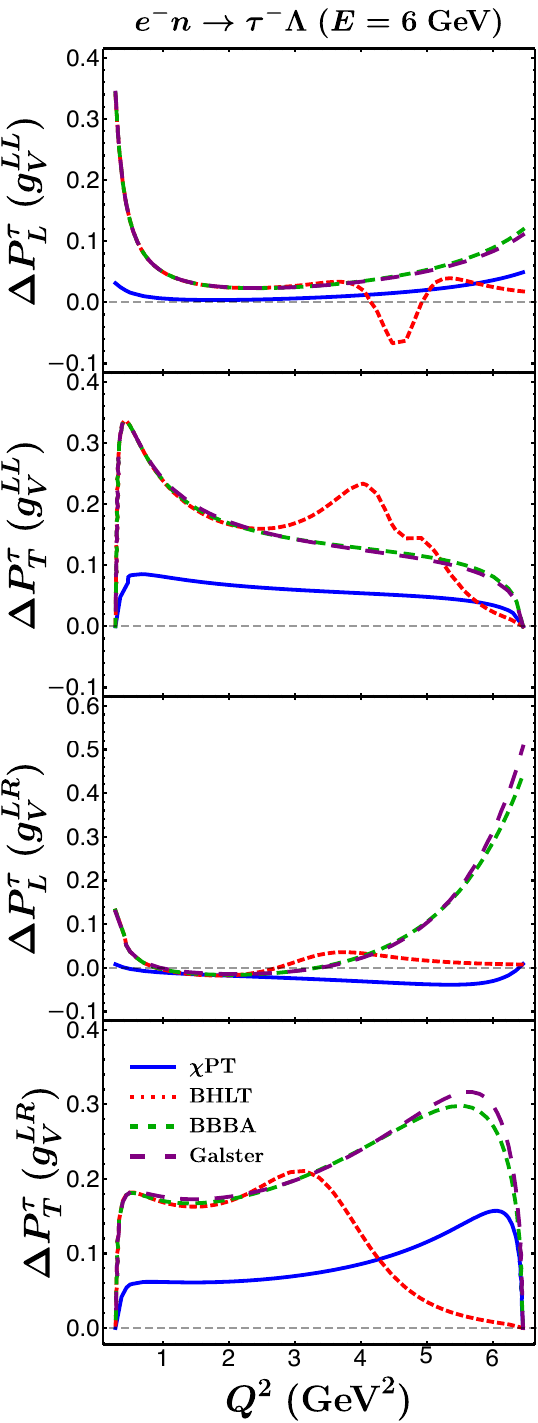}
	\end{minipage}}
	\caption{Effects of the $g_3$ form factor to the tau polarization observables, $\Delta P^{\tau}_{L,T} \equiv P^{\tau}_{L,T} - P^{\tau}_{L,T}(g_3=0)$, for $e^-p \to \tau^-\Sigma^+$ (left) and $e^-n \to \tau^-\Lambda$ (right) at $E=6$~GeV. 
		The $P^{\tau}_{L,T}$ arising from a scalar interaction ($g_S^{LL}$) is independent of the form factors (see Figs.~\ref{fig:PSig} and \ref{fig:PLam}), thus its corresponding $\Delta P^{\tau}_{L,T}$ is zero and not shown.}
	\label{fig:g3Sig1}
\end{figure}
 
\subsection{Effects of $g_3$}
\label{sec:g3}

As discussed in Sec.~\ref{subsec:Cross section}, the hadronic matrix elements for QE scattering can 
also be extracted from analyses of hyperon semileptonic decays.
In the context of these decays involving light leptons ($\ell=e, \mu$), the form factor $g_3$ is often neglected, as its contribution is suppressed by the small lepton mass ratio $m_\ell/m_Y$~\cite{Schlumpf:1994fb, Sasaki:2008ha}. Consequently, some form factor parametrizations derived from this context, such as the QCDSR model~\cite{Zhang:2024ick} employed in this work, set $g_3=0$ by construction.

However, for the tau-lepton final states considered in this work, such suppression is absent, as $m_\tau/m_Y \sim \mathcal{O}(1)$. The neglect of $g_3$ may therefore introduce significant theoretical uncertainties. To gauge the potential magnitude of this effect and assess the validity of using a $g_3=0$ approximation, we systematically investigate the impact of this form factor within the other available parametrizations by comparing calculations with and without its inclusion.

In Fig.~\ref{fig:g3Sig1}, we examine how the $g_3$ form factor affects the polarization observables $P_{L,T}^\tau$. The plots show $Q^2$ dependence for the $e^-p \to \tau^-\Sigma^+$ and $e^-n \to \tau^-\Lambda$ processes, respectively, calculated with a lower beam energy of $E=6$~GeV, where the $P_{T}^\tau$ can be larger (see Figs.~\ref{fig:AvgPSig} and \ref{fig:AvgPLam}).
Regarding the $P_L^\tau$, the contribution from $g_3$ is typically small, except in specific kinematic regimes.  Moderate effects emerge at low $Q^2$ for the $e^-n \to \tau^-\Lambda$ process in the $g_V^{LL}$ scenario (top right), and more prominently as a large enhancement at high $Q^2$ for both processes in the $g_V^{LR}$ case (third row).
In contrast, the $P_T^\tau$ reveals a much richer and more model-dependent sensitivity to $g_3$. For the $e^-p \to \tau^-\Sigma^+$ reaction, all models still show minimal impact for the $g_V^{LL}$ operator. For the $g_V^{LR}$ operator, however, the BBBA and Galster models predict a notable effect from $g_3$. The models predict an even more pronounced dependence on $g_3$ for the $e^-n \to \tau^-\Lambda$ process. 
Here, the $\chi$PT model is the only one that remains largely insensitive to the inclusion of $g_3$, while all others show noticeable deviations.

\begin{figure}[t]
	\centering
	\subfigure{\begin{minipage}{0.238\textwidth}
			\centering
			\includegraphics[width=1.64in]{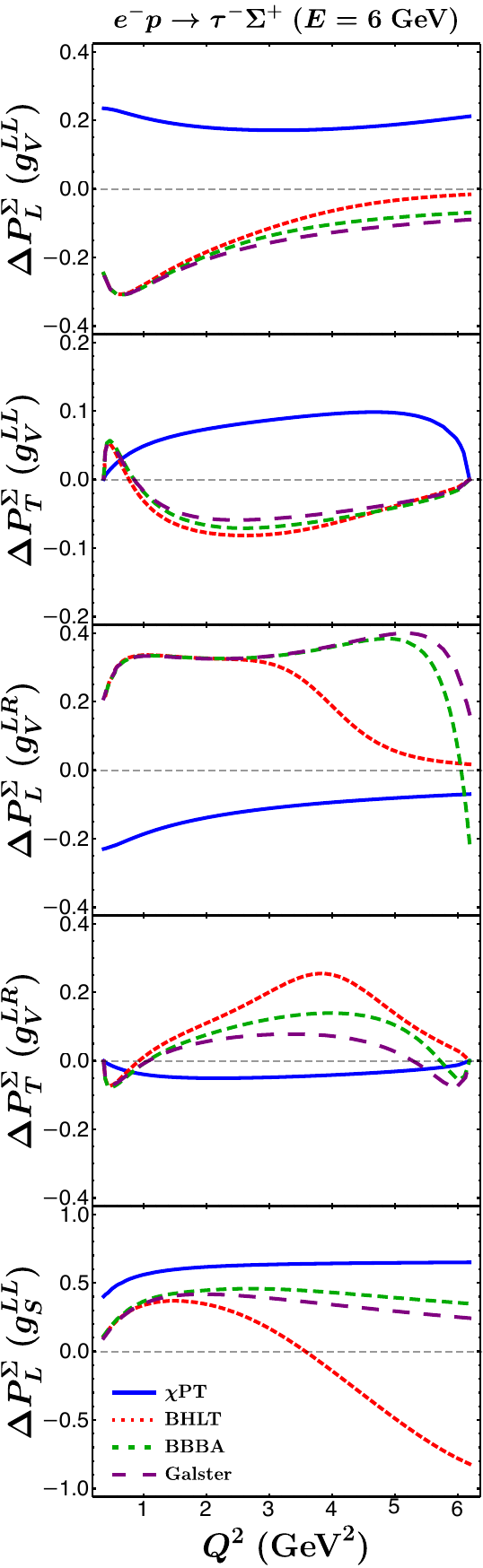}
	\end{minipage}}\hspace{0.018cm}
	\subfigure{\begin{minipage}{0.238\textwidth}
			\includegraphics[width=1.64in]{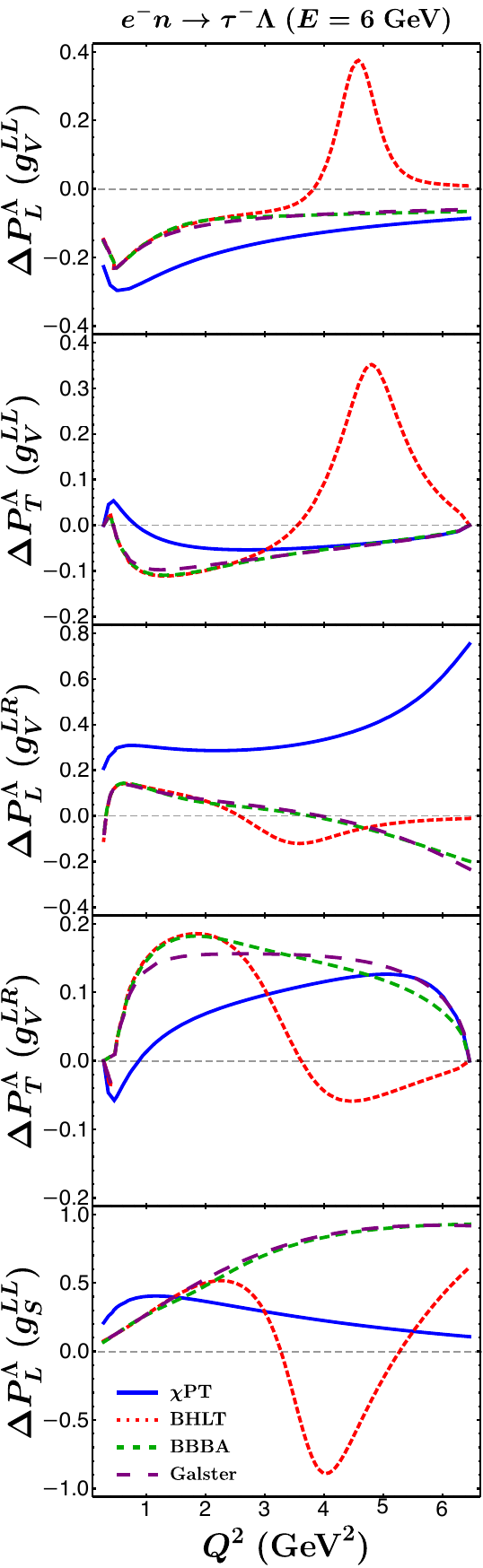}
	\end{minipage}}
	\caption{Effects of the $g_3$ form factor to the hyperon polarization observables, $\Delta P^{Y}_{L,T} \equiv P^{Y}_{L,T} - P^{Y}_{L,T}(g_3=0)$, for the process $e^-p \to \tau^-\Sigma^+$ (left) and  $e^-n \to \tau^-\Lambda$ (right) at $E=6$~GeV. The $P^{Y}_{T}$ arising from a scalar interaction ($g_S^{LL}$) is equal to zero (see Figs.~\ref{fig:PSig} and \ref{fig:PLam}), thus its corresponding $\Delta P^{Y}_{T}$ is not shown.}
	\label{fig:g3Sig2}
\end{figure}

We now extend our analysis to the hyperon polarizations, $P_{L,T}^{\Sigma, \Lambda}$, investigating their sensitivity to the form factor $g_3$ in Fig.~\ref{fig:g3Sig2}. Our examination reveals a clear dichotomy: the impact of $g_3$ is generally modest for the vector operators but becomes dominant for the scalar operator.

For the vector operators $\mathcal{O}_V^{LL}$ and $\mathcal{O}_V^{LR}$, the inclusion of $g_3$ has a limited effect on the hyperon polarizations in most scenarios. For the $\Sigma^+$ final state, the predictions for $P_{L,T}^\Sigma$ are largely insensitive to $g_3$, with the only notable exception being a deviation in $P_L^\Sigma$ at intermediate $Q^2$ for the $g_V^{LR}$ scenario within the BHLT, BBBA, and Galster models.  
A similar overall insensitivity is observed for the $\Lambda$ final state, but with different exceptions: here, the primary effect of $g_3$ is seen in the $P_L^\Lambda$ for the $g_V^{LR}$ operator within the $\chi$PT model, and in both polarization components $P_{L,T}^\Lambda$ for the $g_V^{LL}$ case within the BHLT model.

In contrast, the situation changes dramatically for the scalar operator, $\mathcal{O}_S^{LL}$ (bottom row). Here, the $g_3$ form factor plays a crucial role for both processes. Including $g_3$ completely reshapes the $Q^2$ dependence of the longitudinal polarization ($P_L^{\Sigma, \Lambda}$), often altering its magnitude significantly. 
Such a profound sensitivity indicated that any reliable prediction for scalar NP scenarios is critically dependent on an accurate 
determination of the $g_3$ form factor.

\section{Experimental Prospects}
\label{sec:Prospect}

In this section, we evaluate the experimental prospects for observing the LFV scattering processes $e^-N \to \tau^-Y$. 
Although our theoretical treatment includes polarization observables, the primary challenge is the expected low event rate. 
Given that low statistics may preclude the measurement of polarization asymmetries, we here focus our analysis on the total cross section.

\subsection{Experimental Setup}

To estimate the potential event rates, we consider a specific high-luminosity experimental setup. We select the proposed Solenoidal Large Intensity Device (SoLID) at the 12 GeV CEBAF facility at JLab~\cite{Arrington:2021alx, JeffersonLabSoLID:2022iod}. This choice is motivated by its suitability for studying the sub-threshold $J/\psi$ production from deuteron targets~\cite{Liu:2023htv, K.Hafidi}, where the final state is reconstructed from dilepton decay products. By analogy, we assume that the final-state $Y$ and the visible decay products of the $\tau$ can also be efficiently reconstructed at SoLID. 
Following the setup outlined in Ref.~\cite{Liu:2023htv}, we adopt a 12~GeV electron beam with an intensity of 1.25~$\mu$A on a liquid deuteron target. This corresponds to an integrated luminosity of $\mathcal{L}=1.2\times10^{37}$~cm$^{-2}$s$^{-1}$, which we use to estimate the event rate via the relation $dN/dt = \mathcal{L}\sigma$.

\subsection{Constraints on Wilson Coefficients}

To provide a benchmark for our cross-section calculations, we must determine the current constraints on the relevant WCs. 
The most stringent bounds arise from experimental searches for the cLFV tau decays, namely $\tau^-\to e^- K_S$, $\tau^-\to e^-K^{*0}$, and $\tau^{-} \to e^{-} \pi^{-} K^{+}$. While constraints also exist from high-$p_T$ dilepton invariant mass analysis at the LHC, 
it has been demonstrated that the low-energy $\tau$ decays provide the most restrictive limits~\cite{Angelescu:2020uug} , 
which we therefore focus on.

The current experimental limits at 90\% confidence level on the branching fractions for these decays are:
\begin{align}
	\mathcal{B}(\tau^-\to e^-K_S) &< 2.6\times 10^{-8}~\text{\cite{Belle:2010rxj}}\,, \\
	\mathcal{B}(\tau^-\to e^-K^{*0}) &< 1.9\times 10^{-8}~\text{\cite{Belle:2023ziz}}\,, \\
	\mathcal{B}(\tau^{-} \to e^{-} \pi^{-} K^{+})&<3.1\times10^{-8}~\text{\cite{Belle:2012unr}}\,.
\end{align}
By applying these limits to the decay rate formulas, we derive the upper bounds on the individual WCs. We find that the $\tau^-\to e^-K^{*0}$ decay provides the most restrictive constraint on the vector coefficients:
\begin{align}\label{eq:cons_gV}
	|g^{\alpha}_V|<2.77\times10^{-4} \quad (\alpha=\text{LL, LR, RL, RR})\,. 
\end{align} 
On the other hand, the strongest constraint on the scalar coefficients comes from the $\tau^{-} \to e^{-} \pi^{-} K^{+}$ decay:
\begin{align}\label{eq:cons_gS}
	|g^{\alpha}_S|<4.09\times10^{-4} \quad (\alpha=\text{LL, LR, RL, RR})\,. 
\end{align} 
A detailed discussion of the decay rate calculations can be found in Appendix~\ref{app:constraints} and Ref.~\cite{He:2019iqf}.

\subsection{Event Rate Projections and Discussion}

We now present the projected annual event rates for the SoLID experiment in Table~\ref{table:event}. Our calculations assume a one-year run time, incorporate the 70\% combined branching fraction from dominant $\tau$ decay channels, and adopt an idealized 100\% reconstruction efficiency for all visible final-state particles. The projections are derived for various nucleon form factor models, under the assumption of a single dominant WC set to its maximum allowed value.

\begin{table}[!thpb]
	\renewcommand\arraystretch{1.4} 
	\tabcolsep=0.24cm
	\centering
	\caption{Projected annual event rates ($N/\text{yr}$) for $e^- N \to \tau^- Y$ processes at the proposed SoLID experiment. The rates are calculated for different nucleon form factor models, assuming a single non-zero WC at its maximum allowed value and incorporating a 70\% branching fraction for the dominant $\tau$ decay channels. The rate for the $e^-n\to\tau^- \Sigma^0$ process is not listed, as it is predicted to be half that of $e^-p\to\tau^- \Sigma^+$ (see Table~\ref{tab:SU3}).}
	\begin{tabular}{lccccc}
		\hline\hline
		\multicolumn{6}{c}{$e^-p\to\tau^- \Sigma^+$}  \\ \hline
		Model	& QCDSR & $\chi$PT & BHLT & BBBA & Galster  \\ \hline
		$g^{LL,RR}_V $& 6.73& 0.668&0.0473 &0.0206&0.0211  \\ 
		$g^{LR,RL}_V$&3.90 &1.20 &0.0425 &0.0185 &0.0190  \\
		$g^{\alpha}_S$&183 &2.46 &0.117 &0.00372&0.00389 \\
		\hline \hline
		\multicolumn{6}{c}{$e^-n\to\tau^- \Lambda$}  \\ \hline
		Model	& QCDSR & $\chi$PT & BHLT & BBBA & Galster  \\ \hline
		$g^{LL,RR}_V $&13.3 &2.91 &0.0731 &0.0181 &0.0181  \\ 
		$g^{LR,RL}_V$&4.61 &1.28 &0.0758 &0.0159 &0.0159  \\
		$g^{\alpha}_S$&481 &1.88 &0.136 &0.00208 &0.00209 \\
		\hline \hline
	\end{tabular}
	\label{table:event} 
\end{table}

The final results are summarized in Table~\ref{table:event}, which highlights two primary challenges for an experimental search: low overall event rates and a strong dependence on the choice of hadronic form factor model. Even in the most favorable scenarios (QCDSR, $\chi$PT), the projected yields of a few events per year would preclude any precision measurement of polarization asymmetries at SoLID. A significant advancement is possible, however, at future facilities like JLab's Hall C, which promises both higher luminosities ($10^{38}$ to $10^{39}$~cm$^{-2}$s$^{-1}$) and full polarization capabilities~\cite{Arrington:2021alx}. This combination is crucial; the enhanced luminosity could provide the necessary statistics, while the availability of polarized beams and targets would allow for deeper probes of the new physics. Indeed, if the QCDSR or $\chi$PT scenarios are representative of nature, a high-luminosity experiment at Hall C could render the measurement of polarization asymmetries statistically viable.

Let us conclude this section by making the following comment. 
To fully characterize the new physics in Eq.~\eqref{eq:Leff}, a global analysis that moves beyond the single-operator framework is ultimately required. This necessitates a program of measurements sensitive to the interference between different operators. The polarization components ($P_i$) of the final-state $\tau$ and $Y$, though statistically demanding, are particularly well-suited for this task, 
since they are sensitive to different linear combinations of operators. 
Furthermore, doubly-polarized scattering experiments, employing both polarized beams and targets, offer another powerful probe. 
As highlighted in Ref.~\cite{Lai:2022ekw}, this technique directly targets the Dirac structure of the new physics, 
providing complementary information needed to construct a complete picture of the cLFV interaction. 

\section{Conclusion}
\label{sec:conclusion}

In this work, we have performed a comprehensive study of the lepton-flavor-violating quasi-elastic scattering processes 
$e^- N \to \tau^- Y$. 
Working in framework of a general low-energy effective Lagrangian given by Eq.~\eqref{eq:Leff}, we have analyzed the 
the polarization observables of the final-state $\tau$ lepton and hyperon $Y$.

A central finding of our analysis is the profound sensitivity of all observables to the choice of hadronic form factor parametrization. We have compared five distinct models (BBBA, Galster, BHLT, QCDSR, and $\chi$PT) and found that the predicted cross sections span several orders of magnitude, with the QCDSR and $\chi$PT models yielding rates significantly larger than the others. The polarization observables show a similar, dramatic sensitivity; the shape and even the sign of the transverse polarization $P^Y_T$ can differ completely between form factor schemes. This strong model dependence constitutes a major theoretical uncertainty that must be addressed for a robust interpretation of future experimental data.

We have then demonstrated that the final-state averaged polarization observables themselves offer a powerful means to resolve these theoretical ambiguities. For instance, in vector NP scenarios, the predicted hyperon polarizations, $\langle P_{L,T}^\Sigma \rangle$ and $\langle P_L^\Lambda \rangle$, 
could separate the form factor models into distinct classes. 
Building on this discriminatory power, we have outlined a strategy whereby a combined measurement of 
$\langle P_L^\tau \rangle$ and $\langle P_T^\Sigma \rangle$ would then allow for the unique identification 
of the underlying vector operator.

We have also investigated the impact of the form factor, $g_3$, on the QE scattering processes involving the heavy tau lepton. While its effect on the vector-current-induced polarization observables is generally modest, its inclusion dramatically alters predictions for the scalar-current scenarios. This underscores the necessity of a precise determination of $g_3$ in the spacelike region for any reliable interpretation of potential signals from scalar cLFV interactions.

Finally, we have assessed the discovery potential of the QE processes at the proposed SoLID experiment at JLab. Using current constraints on the WCs from the cLFV $\tau$ decays, we have found that annual event rates span several orders of magnitude---from negligible levels to several hundred 
events---a range dictated majorly by the underlying form factor model. While a simple discovery through event counting might be possible in the most optimistic scenarios (\emph{e.g.}, QCDSR or $\chi$PT), precision measurements of polarization observables would likely require the higher luminosities envisioned for future facilities.

\section*{Acknowledgments}
This work is supported by the National Natural Science Foundation of China under Grants No. 12135006 and No. 12275067, 
the Natural Science Foundation of Henan Province under Grant No. 242300421390, 
the Science and Technology R\&D Program Joint Fund Project of Henan Province under Grant No. 225200810030, 
the Science and Technology Innovation Leading Talent Support Program of Henan Province under Grant No. 254000510039, 
as well as the National Key R\&D Program of China under Grant No. 2023YFA1606000.

\appendix

\section{Hadronic and leptonic matrix elements}
\label{app:matrix ele}

In this appendix, we collect the definitions of the hadronic and leptonic matrix elements that compose the 
amplitude in Eq.~\eqref{eq:amplitude}, within the framework of the effective Lagrangian $\mathcal{L}_{\text{eff}}$.

The hadronic matrix elements $H^{L,R}_a$ are defined as follows:
\begin{align}
	H^L_S=&\frac{1}{2}\langle Y(p^\prime,t)|\big[\left(g^{LL}_S+g^{LR}_S\right)\bar{s}d \nonumber \\[0.02cm]
	&+\left(g^{LR}_S-g^{LL}_S\right)\bar{s}\gamma^5d\big]|N(p,s)\rangle\,, \nonumber \\[0.02cm]
	H^R_S=&\frac{1}{2}\langle Y(p^\prime,t)|\big[\left(g^{RR}_S+g^{RL}_S\right)\bar{s}d \nonumber \\[0.02cm]
	&+\left(g^{RR}_S-g^{RL}_S\right)\bar{s}\gamma^5d\big]|N(p,s)\rangle\,, \nonumber \\[0.02cm]
	H^{L\mu}_V=&\frac{1}{2}\langle Y(p^\prime,t)|\big[\left(g^{LL}_V+g^{LR}_V\right)\bar{s}\gamma^{\mu}d \nonumber \\[0.02cm] &+\left(g^{LR}_V-g^{LL}_V\right)\bar{s}\gamma^{\mu}\gamma^5d\big]|N(p,s)\rangle\,, \nonumber  \\[0.02cm]
	H^{R\mu}_V=&\frac{1}{2}\langle Y(p^\prime,t)|\big[\left(g^{RR}_V+g^{RL}_V\right)\bar{s}\gamma^{\mu}d\nonumber\\[0.02cm]
	&+\left(g^{RR}_V-g^{RL}_V\right)\bar{s}\gamma^{\mu}\gamma^5d\big]|N(p,s)\rangle\,, \nonumber  \\[0.02cm]
	H^{L\mu\nu}_T=&g^{LL}_T\langle Y(p^\prime,t)|\bar{s}\sigma^{\mu\nu} d |N(p,s)\rangle\,,\nonumber  \\[0.02cm]
	H^{R\mu\nu}_T=&g^{RR}_T\langle Y(p^\prime,t)|\bar{s}\sigma^{\mu\nu} d |N(p,s)\rangle\,.
\end{align}
The corresponding leptonic matrix elements $L^{L,R}_a$ are given by:
\begin{align}
	L^L_S&=\bar{u}(k^{\prime},\lambda)P_Lu(k,r)\,, \nonumber  \\[0.02cm]
	L^R_S&=\bar{u}(k^{\prime},\lambda)P_Ru(k,r)\,, \nonumber  \\[0.02cm]
	L^L_{V\mu}&=\bar{u}(k^{\prime},\lambda)\gamma_{\mu} P_Lu(k,r)\,,\nonumber  \\[0.02cm]
	L^R_{V\mu}&=\bar{u}(k^{\prime},\lambda)\gamma_{\mu}  P_Ru(k,r)\,,\nonumber  \\[0.02cm]
	L^L_{T\mu\nu}&=\bar{u}(k^{\prime},\lambda)\sigma_{\mu\nu} P_Lu(k,r)\,,\nonumber  \\[0.02cm]
	L^R_{T\mu\nu}&=\bar{u}(k^{\prime},\lambda)\sigma_{\mu\nu} P_Ru(k,r)\,.
\end{align}
Here, $r$ and $s$ ($\lambda$ and $t$) denote the helicities of the initial (final) lepton and baryon, respectively. 
The corresponding four-momenta are denoted by $k$ and $p$ ($k'$ and $p'$).

\section{Parameterizations of the form factors $G_{E,M}^{p,n}$ }
\label{app:FFparam}

We summarize below the parameterizations of the nucleon electromagnetic form factors employed in this study. 
\subsection{Galster Parameterization}

The Galster parameterization~\cite{Galster:1971kv} is expressed as
\begin{align}
	G_E^p(q^2) &= \left( 1 - \frac{q^2}{M_V^2} \right)^{-2}\,, \nonumber \\[0.02cm]
	\frac{G_M^p(q^2)}{\mu_p} &= \frac{G_M^n(q^2)}{\mu_n} = G_E^p(q^2)\,, \nonumber \\[0.02cm]
	G_E^n(q^2) &= - \frac{\mu_n \tau}{1 + \lambda_n \tau} G_E^p(q^2)\,,
\end{align}
where $M_V = 0.843~\mathrm{GeV}$, $\mu_p = 2.7928$, $\mu_n = -1.9130$, $\lambda_n = 5.6$, and $\tau =-q^2/(4m_N^2)$ with $m_N$ the nucleon mass.

\subsection{BBBA Parameterization}

The BBBA parameterization~\cite{Bradford:2006yz} adopts a rational functional form:
\begin{align}
	G_E^p(q^2)&=\frac{1-0.0578\tau}{1+11.1\tau+13.6\tau^2+33.0\tau^3}\,, \nonumber \\[0.02cm]
	\frac{G_M^p(q^2)}{\mu_p}&=\frac{1+0.15\tau}{1+11.1\tau+19.6\tau^2+7.54\tau^3}\,, \nonumber \\[0.02cm]
	G_E^n(q^2)&=\frac{1.25\tau+1.30\tau^2}{1-9.86\tau+305\tau^2-758\tau^3+802\tau^4}\,, \nonumber \\[0.02cm]
	\frac{G_M^n(q^2)}{\mu_n}&=\frac{1+1.81\tau}{1+14.1\tau+20.7\tau^2+68.7\tau^3}\,.
\end{align}

\subsection{BHLT Parameterization}

The BHLT parameterization~\cite{Borah:2020gte} employs a model-independent expansion in the conformal variable $z(q^2)$, defined as
\begin{equation}
	z(q^2) = \frac{\sqrt{t_{\mathrm{cut}} - q^2} - \sqrt{t_{\mathrm{cut}} - t_0}}{\sqrt{t_{\mathrm{cut}} - q^2} + \sqrt{t_{\mathrm{cut}} - t_0}},
\end{equation}
with $t_{\mathrm{cut}} = 4 m_\pi^2$ and $t_0 = -0.21~\mathrm{GeV}^2$. The form factors are expanded as
\begin{align}
	G_E^{p,n}(q^2) &= \sum_{k=0}^{k_{\mathrm{max}}} a_k\, z(q^2)^k\,, \nonumber \\[0.02cm]
	G_M^{p,n}(q^2) &= G_M^{p,n}(0) \sum_{k=0}^{k_{\mathrm{max}}} b_k\, z(q^2)^k\,,
\end{align}
where the coefficients $a_1$–$a_4$ and $b_1$–$b_4$ are determined from fits provided in Ref.~\cite{Borah:2020gte}.

The remaining coefficients, namely $a_0$ and $a_5$--$a_8$, are fixed by imposing the normalization conditions at $q^2 = 0$,
\begin{align}
	G_E^p(0) \!=\! 1\,, \ G_E^n(0) \!=\! 0\,, \ G_M^p(0)\! =\! \mu_p\,, \ G_M^n(0)\! =\! \mu_n\,,
\end{align}
together with a set of sum rules derived in Refs.~\cite{Borah:2020gte,Lee:2015jqa},
\begin{align}\label{eq:sumrules}
	&  \sum_{k=0}^8 a_k = 0\,,\quad   \sum_{k=1}^8 k a_k = 0\,, \nonumber \\[0.02cm]
	&  \sum_{k=2}^8 k(k-1) a_k = 0\,,\quad  \sum_{k=3}^8 k(k-1)(k-2) a_k = 0\,.
\end{align}
An analogous procedure can be applied to determine the coefficients $b_0$ and $b_5$--$b_8$ for the magnetic form factors.

\section{Form factors for $Y\to N$ transitions}
\label{app:form factors}
      
As discussed in Sec.~\ref{subsec:Cross section}, the hadronic matrix elements for $N\to Y$ ($d\to s$) transitions 
can also be obtained via the complex conjugate of the corresponding $Y\to N$ ($s\to d$) matrix elements. 
For completeness, we list below the relevant vector and axial-vector matrix elements for the $Y\to N$ ($s\to d$) transitions, 
\begin{align}
	&\quad \langle N(p) |\bar{d}\gamma_{\mu}s|Y(p^{\prime})\rangle \nonumber \\
	&=\bar{u}(p)\left[\gamma_\mu f^\prime_1(q^2)+i\sigma_{\mu\nu}\frac{q^\nu}{m_Y}f^\prime_2(q^2)
	+\frac{q_\mu}{m_Y}f^\prime_3(q^2)\right]u(p^{\prime})\,, \label{eq:Decay_V}\\
	&\quad	\langle N(p) |\bar{d}\gamma_{\mu}\gamma_5 s|Y(p^{\prime})\rangle \nonumber \\
	&=\bar{u}(p)\left[\gamma_\mu g^\prime_1(q^2)
	+i\sigma_{\mu\nu}\frac{q^\nu}{m_Y}g^\prime_2(q^2)+\frac{q_\mu}{m_Y}g^\prime_3(q^2)\right]
	\gamma_5u(p^{\prime})\,, \label{eq:Decay_A} 
\end{align}
where $q=p^\prime-p$, and $f^\prime_i(q^2)$ and $g^\prime_i(q^2)$ ($i$=1, 2, 3) denote the vector and axial-vector transition form factors, respectively.

\begin{table}[tp]
	\renewcommand*{\arraystretch}{1.5}
	\tabcolsep=0.4cm
	\centering
	\caption{Values of $\mathcal{V}_{NY}$ and $\mathcal{A}_{NY}$ for the transitions $NY = n\Lambda, n\Sigma^0, p\Sigma^+$. The constants $D$ and $F$ are taken from the lowest-order chiral Lagrangian.}
	\begin{tabular}{lccc}
		\hline \hline
		$NY$ & $n\Lambda$ & $n\Sigma^0$ & $p\Sigma^+$ \\
		\hline		
		$\mathcal{V}_{NY}$ & $-\sqrt{\frac{3}{2}}$ & $\frac{1}{\sqrt{2}}$  & $-1$ \\       
		$\mathcal{A}_{NY}$ & $\frac{-1}{\sqrt{6}}(D+3F)$ & $\frac{-1}{\sqrt{2}}(D-F)$ & $D-F$          \\                                                   
		\hline \hline		                                                                                                
	\end{tabular}	
	\label{tab:Chiral}
\end{table}

In the framework of $\chi$PT,  the matrix elements take the following simplified forms~\cite{Tandean:2019tkm}:
\begin{align}\label{eq:chi_V}
	\langle N(p)|\bar{d}\gamma_{\mu}s|Y(p^{\prime})\rangle&=\mathcal{V}_{NY}\bar{u}(p)\gamma_{\mu}u(p^{\prime})\,,
\end{align}
and 
\begin{align}\label{eq:chi_A}
&\langle N(p) |\bar{d}\gamma_{\mu}\gamma_5 s|Y(p^{\prime})\rangle \nonumber \\ 
&=\mathcal{A}_{NY}\bar{u}(p)\left(\gamma_{\mu}-\frac{m_N+m_Y}{m^2_{K^0}-q^2}q_\mu \right)\gamma_5u(p^{\prime})\,,
\end{align}
where $m_{K^0}$ is the mass of the neutral kaon, and $\mathcal{V}_{NY}$ and $\mathcal{A}_{NY}$ 
are constants whose values are summarized in Table~\ref{tab:Chiral}. 
These couplings are derived from the lowest-order chiral Lagrangian, 
with the axial couplings $D = 0.81$ and $F = 0.47$ determined from fits to 
baryon semileptonic decay data~\cite{ParticleDataGroup:2024cfk}.
By comparing the matrix elements in Eqs.~\eqref{eq:chi_V} and \eqref{eq:chi_A} with our 
decomposition conventions in Eqs.~\eqref{eq:Decay_V} and \eqref{eq:Decay_A}, it is straightforward 
to verify that $f_2^\prime = f_3^\prime = g_2^\prime = 0$.

The QCDSR study~\cite{Zhang:2024ick} employs the same Lorentz decomposition of the vector and axial–vector matrix elements as in Eqs.~\eqref{eq:Decay_V} and \eqref{eq:Decay_A} for the charged–current $Y\to N$ ($s\to u$) transitions. 
We extend this scheme to the neutral-current $Y \to N$ ($s \to d$) case by invoking isospin symmetry. Under this assumption, one finds that $f_3^\prime = g_2^\prime = g_3^\prime = 0$. For the nonvanishing form factors $f_i^\prime(q^2)$ and $g_i^\prime(q^2)$, the QCDSR employs a $z$-expansion parametrization:
\begin{align}
	f_i^\prime(q^2) &= \frac{f^\prime_i(0)}{1 - q^2/m_{\text{pole}}^2} \left[1 + a_1\left(z(q^2) - z(0)\right)\right]\,, \\[0.02cm]
	z(q^2) &= \frac{\sqrt{t_+ - q^2} - \sqrt{t_+ - t_0}}{\sqrt{t_+ - q^2} + \sqrt{t_+ - t_0}}\,,
\end{align}
where $t_\pm = (m_Y \pm m_N)^2$, $t_0 = t_+ - \sqrt{t_+ - t_-}\sqrt{t_+ - t_{\text{min}}}$, and $t_{\text{min}} = -0.05~\mathrm{GeV}^2$. 
It adopts $m_{\rm pole}=0.892~\mathrm{GeV}$ for $f_1'$ and  
$m_{\rm pole}=1.27~\mathrm{GeV}$ for $g_1'$, and neglects the $q^2$ dependence of $f_2'$.
For the numerical values of $a_1$, $f^\prime_1(0)$, and $g^\prime_1(0)$ in each transition, we refer the reader to Ref.~\cite{Zhang:2024ick}.

\section{$|\Delta S|=1$ leptonic and semileptonic $\tau$ decays}
\label{app:constraints}

This section details the formalism for calculating the rates of lepton flavor violating $\tau$ decays. The calculations are based on the effective Lagrangian, $\mathcal{L}_{\text{eff}}$, presented in Eq.~\eqref{eq:Leff}. Our approach follows the general framework of Ref.~\cite{He:2019iqf}, from which we also adapt the necessary hadronic matrix elements.

\subsection*{1. Two-Body Decays}

\subsubsection*{a. Pseudoscalar Meson Final States ($\tau \to \ell P$)}

For the decay of a $\tau$ lepton into a charged lepton $\ell$ and a pseudoscalar meson $P$, the general amplitude is given by
\begin{align}
	\mathcal{M}_{\tau\to\ell P}=i \bar{u}_{\ell}\left(\mathcal{S}^{\ell}_P+\gamma_5 \mathcal{P}^{\ell}_P\right)u_{\tau}\,.
\end{align}
The corresponding decay rate is then
\begin{align}
	\Gamma_{\tau \to \ell P} =& \frac{\mathcal{K}^{1/2}(m_{\tau}^2, m_{\ell}^2, m_{P}^2)}{16\pi m_{\tau}^3} \Big\{ \left[ (m_{\tau} + m_{\ell})^2 - m_{P}^2 \right] \left| \mathcal{S}_{P}^{\ell} \right|^2\nonumber\\
	& + \left[ (m_{\tau} - m_{\ell})^2 - m_{P}^2 \right] \left| \mathcal{P}_{P}^{\ell} \right|^2 \Big\}\,,
\end{align}		
where $\mathcal{K}(x,y,z)=(x-y-z)^2-4yz$ is the K\"all\'en function.

To obtain the $\mathcal{S}^{\ell}_P$ and $\mathcal{P}^{\ell}_P$ for $\tau^-\to \ell^- K_S$ decay, 
we require the relevant hadronic matrix elements:
\begin{align}
	\langle K^0 | \bar{d} \gamma^\mu \gamma_5 s | 0 \rangle &= \langle \bar{K}^0 | \bar{s} \gamma^\mu \gamma_5 d | 0 \rangle = i f_K p_K^\mu\,, \\[0.02cm]
	\langle K^0 | \bar{d} \gamma_5 s | 0 \rangle &= \langle \bar{K}^0 | \bar{s} \gamma_5 d | 0 \rangle = i B_0 f_K \,,
\end{align}
where $f_K$ is the kaon decay constant and $B_0=m^2_{K^0}/(m_d+m_s)$. Using the approximation $K_S \approx (K^0-\bar{K}^0)/\sqrt{2}$, 
we obtain
\begin{align}
	\mathcal{S}_{K_S}^{\ell} &= \frac{f_KG_F}{2}\left[(m_{\tau} - m_{\ell})\tilde{V}_{\ell\tau} + B_0 \tilde{S}_{\ell\tau}\right], \label{eq:SKS}\\[0.02cm]
	\mathcal{P}_{K_S}^{\ell} &= \frac{f_KG_F}{2}\left[-(m_{\tau} + m_{\ell})\tilde{A}_{\ell\tau} + B_0 \tilde{P}_{\ell\tau}\right], \label{eq:PKS}
\end{align}
with 
\begin{align}
	\tilde{V}_{\ell\tau} &= -g^{LL}_V+g^{LR}_V-g^{RL}_V+g^{RR}_V\,,\\[0.02cm]
	\tilde{A}_{\ell\tau} &= g^{LL}_V-g^{LR}_V-g^{RL}_V+g^{RR}_V\,, \\[0.02cm]
	\tilde{S}_{\ell\tau} &= g^{LL}_S-g^{LR}_S+g^{RL}_S-g^{RR}_S\,,\\[0.02cm]
	\tilde{P}_{\ell\tau} &= g^{LL}_S-g^{LR}_S-g^{RL}_S+g^{RR}_S\,. 
\end{align}
It should be noted that the derivation of Eqs.~\eqref{eq:SKS} and \eqref{eq:PKS} only includes contributions from the $\tau^-\to \ell^- K^0$ decay, as this is the specific process relevant to the QE scattering analyses in this work.

\subsubsection*{b. Vector Meson Final States ($\tau \to \ell V$)}

Similarly, for decays into a vector meson $V$, the amplitude is
\begin{align}
	\mathcal{M}_{\tau\to\ell V}=\bar{u}_{\ell}\slashed{\varepsilon}_V \left(\mathcal{V}^{\ell}_V+\gamma_5 \mathcal{A}^{\ell}_V\right)u_{\tau}\,,
\end{align}
where $\varepsilon_V$ is the polarization vector of the meson. This leads to the decay rate
\begin{align}
	\Gamma_{\tau \to \ell V}\! = & \frac{\mathcal{K}^{1/2}(m_{\tau}^2, m_{\ell}^2, m_{V}^2)}{16\pi m_{\tau}^3 m_{V}^2}\nonumber \\
	&\!\times\! \Big\{ [\tilde{K}(m_{\tau}^2, m_{\ell}^2, m_{V}^2) - 6m_{\tau}m_{\ell}m_{V}^2] |\mathcal{V}_{V}^{\ell}|^2 \nonumber\\
	&\!+\! [\tilde{K}(m_{\tau}^2, m_{\ell}^2, m_{V}^2)\! +\! 6m_{\tau}m_{\ell}m_{V}^2] |\mathcal{A}_{V}^{\ell}|^2 \Big\},
\end{align}
where the kinematic function is $\tilde{K}(x, y, z) = (x - y)^2 + (x + y)z - 2z^2$. For the specific decay $\tau^-\to \ell^-K^{*0}$, the necessary mesonic matrix element is
\begin{align}
	\langle K^{*0} | \bar{d} \gamma^\mu s | 0 \rangle  = \varepsilon_{K^*}^\mu f_{K^*} m_{K^*}\,,
\end{align}
where $f_{K^*}$ is the decay constant of the $K^*$. This yields the following relations
\begin{align}
	\mathcal{V}_{K^{*0}}^{\ell}\! =\! \frac{G_F f_{K^*} m_{K^*}}{\sqrt{2}} V_{\ell \tau}\,, \quad 
	\mathcal{A}_{K^{*0}}^{\ell}\! =\! \frac{G_F f_{K^*} m_{K^*}}{\sqrt{2}} A_{\ell \tau}\,,\nonumber
\end{align}
where $V_{\ell \tau}$ and $A_{\ell \tau}$ are combinations of the WCs defined as
\begin{align}
	V_{\ell \tau}&=g^{LL}_V+g^{LR}_V+g^{RL}_V+g^{RR}_V\,, \\[0.02cm]
	A_{\ell \tau}&=-g^{LL}_V-g^{LR}_V+g^{RL}_V+g^{RR}_V\,. 
\end{align}

\vspace{0.1cm}

\subsection*{2. Three-Body Decay: $\tau^{-} \to \ell^{-} \pi^{-} K^{+}$}

For the three-body decay, the hadronic matrix elements are parametrized by form factors $f_0, f_+$, and $f_-$, 
which depend on $q^2=(p_\pi+p_K)^2$:
\begin{align}
	\langle \pi^{-} K^{+} | \bar{d} \gamma^{\mu} s | 0 \rangle & = f_{+} (p_{\pi} - p_{K})^{\mu} - f_{-} q^{\mu}\,,\\[0.02cm]
	\langle \pi^{-} K^{+} | \bar{d} s | 0 \rangle &=  \tilde{B}_{0} f_{0}\,.
\end{align}
Here, $\Delta_{K\pi}^2=m^2_{K^+}-m^2_{\pi^+}$ and $\tilde{B}_0=\Delta_{K\pi}^2/(m_s-m_d)$. The form factors are related by $f_{-} = (f_{0} - f_{+}) \Delta_{K\pi}^2/q^2$. 

The decay amplitude takes the form
\begin{align}
	\mathcal{M}_{\tau \to \ell \pi^{-} K^{+}} = \bar{u}_{\ell} \left( \mathcal{S}_{\pi^{-} K^{+}}^{\ell} + \gamma_{5} \mathcal{P}_{\pi^{-} K^{+}}^{\ell} \right) u_{\tau}\,,
\end{align}
where the scalar and pseudoscalar coefficients are given by
\begin{align}
	\mathcal{S}^{\ell}_{\pi^{-}K^{+}} =&\frac{G_F}{\sqrt{2}}\Big\{ \left[-2f_{+} \slashed{p}_{K} + (f_{+} - f_{-})(m_{\tau} - m_{\ell})\right] V_{\ell\tau}\nonumber \\
	&\qquad + \tilde{B}_{0}f_{0}S_{\ell\tau}\Big\}\,, \\
	\mathcal{P}^{\ell}_{\pi^{-}K^{+}}  =&\frac{G_F}{\sqrt{2}}\Big\{ \left[2f_{+} \slashed{p}_{K} - (f_{+} - f_{-})(m_{\tau} + m_{\ell})\right] A_{\ell\tau}\nonumber\\
	&\qquad  + \tilde{B}_{0}f_{0}P_{\ell\tau}\Big\}\,. 
\end{align}
The relevant combinations of Wilson coefficients are
\begin{align}
	S_{\ell\tau}&=g^{LL}_S+g^{LR}_S+g^{RL}_S+g^{RR}_S\,, \\[0.02cm]
	P_{\ell\tau}&=g^{LL}_S+g^{LR}_S-g^{RL}_S-g^{RR}_S\,. 
\end{align}		
From this amplitude, one can derive the differential decay rate with respect to $\hat{s} = q^2$, 
which is given by~\cite{He:2019iqf} 		
\begin{align}
	&\frac{d\Gamma_{\tau \to \ell\pi^{-}K^{+}}}{d\hat{s}} = \frac{\lambda_{\tau\ell}^{1/2} \lambda_{\pi^{+}K^{+}}^{1/2} |f_0|^2G_F^2}{32\pi^3 m_{\tau}^3} \nonumber \\
	\times &\Bigg\{ \left[ \lambda_{\pi^{+}K^{+}} |f_+|^2 \frac{\lambda_{\tau\ell} + 3\hat{\sigma}_-\hat{s}}{3|f_0|^2 \hat{s}^3} + \Delta_{K\pi}^4 \frac{\lambda_{\tau\ell} + \hat{\sigma}_+\hat{s}}{\hat{s}^3} \right] \frac{|V_{\ell\tau}|^2}{16} \nonumber \\
	&+  \left[ \lambda_{\pi^{+}K^{+}} |f_+|^2 \frac{\lambda_{\tau\ell} + 3\hat{\sigma}_+\hat{s}}{3|f_0|^2 \hat{s}^3} + \Delta_{K\pi}^4 \frac{\lambda_{\tau\ell} + \hat{\sigma}_-\hat{s}}{\hat{s}^3} \right] \frac{|A_{\ell\tau}|^2}{16} \nonumber \\
	&+  \frac{\Delta_{K\pi}^2 \tilde{B}_0}{8\hat{s}^2} \text{Re}(\hat{\mu}_+\hat{\sigma}_- A_{\ell\tau}^* P_{\ell\tau} - \hat{\mu}_-\hat{\sigma}_+ V_{\ell\tau}^* S_{\ell\tau}) \nonumber \\
	&+  \frac{\tilde{B}_0^2}{16\hat{s}} (\hat{\sigma}_+ |S_{\ell\tau}|^2 + \hat{\sigma}_- |P_{\ell\tau}|^2) \Bigg\}\,.
\end{align}		
Note that the kinematic variables are defined as
\begin{align}
	\lambda_{XY} &= \mathcal{K}(m_X^2, m_Y^2, \hat{s})\,, \qquad
	\hat{\sigma}_{\pm} = \hat{\mu}_{\pm}^2 - \hat{s}\,,
\end{align}		
with $\hat{\mu}_{\pm} = m_{\tau} \pm m_{\ell}$.		
		
\bibliographystyle{apsrev4-2}
\bibliography{reference}

\end{document}